\documentclass[nohyper,notoc]{JHEP}

\usepackage{amsmath2000,euscript,array,amssymb,cite} 
\newcommand{\startappendix}{
\setcounter{section}{0}
\renewcommand{\thesection}{\Alph{section}}}
\newcommand{\Appendix}[1]{
\refstepcounter{section}
\begin{flushleft}
{\large\bf Appendix \thesection: #1}
\end{flushleft}}

\def\a0{\phantom{-}0}
\def\ed{\end{document}}
\def\aD{{\dot\alpha}}
\def\bD{{\dot\beta}}
\def\gD{{\dot\gamma}}
\def\dD{{\dot\delta}}
\def\CM{{{\cal M}}}
\def\N{{\cal N}}
\def\tr{{\rm tr}}

\def\sst{\scriptscriptstyle}

\def\SU{\text{SU}}
\def\SO{\text{SO}}
\def\U{\text{U}}

\def\AA{{\boldsymbol A}}
\def\BB{{\boldsymbol B}}

\def\UU{{\boldsymbol Y}}

\newcommand{\BOmega}{{\boldsymbol{\Omega}}}

\newcommand{\BDelta}{{\boldsymbol{\Delta}}}
\newcommand{\Bomega}{{\boldsymbol{\omega}}}
\newcommand{\BSigma}{{\boldsymbol{\Sigma}}}

\newcommand{\Bchi}{{\boldsymbol{\chi}}}
\newcommand{\Bphi}{{\boldsymbol{\phi}}}
\def\Dbarslash{\,\,{\raise.15ex\hbox{/}\mkern-12mu {\bar\D}}}
\def\Dslash{\,\,{\raise.15ex\hbox{/}\mkern-12mu \D}}
\def\delslash{\,\,{\raise.15ex\hbox{/}\mkern-9mu \partial}}
\def\delbarslash{\,\,{\raise.15ex\hbox{/}\mkern-9mu {\bar\partial}}}

\def\ms{{\mathfrak M}}

\def\Z{{\EuScript Z}}

\def\Q{{\cal Q}}

\newcommand{\PD}[2]{\frac{\partial #1}{\partial #2}}
\newcommand{\MAT}[1]{\begin{pmatrix} #1\end{pmatrix}}
\newcommand{\EQ}[1]{\begin{equation} #1 \end{equation}}
\newcommand{\AL}[1]{\begin{subequations}\begin{align} #1
\end{align}\end{subequations}}
\newcommand{\ALT}[2]{\begin{alignat}{#1} #2
\end{alignat}}
\newcommand{\SP}[1]{\begin{equation}\begin{split} #1 \end{split}\end{equation}}

\title{Testing Seiberg-Witten Theory to All Orders in the Instanton Expansion}
\author{Timothy J.~Hollowood\\
Department of Physics, University of Wales Swansea,
Swansea, SA2 8PP, UK\\
E-mail: {\tt t.hollowood@swan.ac.uk}
}

\abstract{In the context of softly-broken $\N=4$ to $\N=2$
supersymmetric $\SU(N)$ gauge theory, we calculate using
semi-classical instanton methods, the lowest order 
non-trivial terms in the mass expansion of the prepotential for all
instanton number. We find exact agreement with 
Seiberg-Witten theory and thereby achieve  
the most powerful test yet of this theory. We also
calculate the one- and two-instanton contributions completely and also
find consistency with Seiberg-Witten theory.
Our approach relies on the fact that the instanton calculus admits a nilpotent
fermionic symmetry, or BRST operator, whose existence implies that the 
integrals over the instanton moduli space, which give the coefficients
of the prepotential, localize on the space of resolved point-like
instantons or what we call ``topicons''.
}


\preprint{{\tt hep-th/0202197}\\SWAT-***}

\begin{document}

\section{Introduction}

The remarkable theory of Seiberg and Witten \cite{SeibWitt} determines
the low-energy behaviour of $\N=2$ supersymmetric gauge theories exactly. In
principle the low-energy effective action can also be calculated from
first principles via conventional semi-classical methods using
instantons. This leads to the idea of testing Seiberg-Witten theory by
calculating the instanton effects and comparing these expressions with
those extracted from the Seiberg-Witten curve. This idea was pursued
most successfully to date in
Refs.~\cite{MO-I,MO-II,DKMn4} at the one- and two-instanton level for the
theory with gauge
group $\SU(2)$ and Ref.~\cite{KMS} at the one-instanton level in $\SU(N)$.
The ultimate goal of this program is to provide an instanton-based
``proof'' of Seiberg-Witten theory by calculating instanton effects to
all orders in the instanton charge. This paper
provides the first test of Seiberg-Witten theory to all instanton number and
the ultimate goal just articulated suddenly looks feasible. We should
mention that there are other semi-classical tests of Seiberg-Witten theory
based on matching the ``monodromies'' of the central charges 
to the semi-classical spectrum of dyons \cite{Hollowood:1998pp}. 

In \cite{local} we described a new technique for calculating the
instanton contributions to the prepotential of $\N=2$ supersymmetric
gauge theory. In these theories there is
an adjoint-valued VEV. Derrick's Theorem \cite{Derrick} implies that
the action of an instanton can always be lowered by shrinking its
size. As a consequence in the presence of a VEV instantons are no
longer solutions of the equations-of-motion. The way to implement the
semi-classical approximation in these circumstances was elucidated by 
Affleck \cite{Affleck,ADS} (see also the in-depth discussion in
Refs.~\cite{MO-I,MY}) leading to the concept of a {\it constrained\/}
instanton. To leading order in the semi-classical expansion the
constrained nature of the instanton manifests itself as a non-trivial
potential, or {\it instanton effective action\/}, on the moduli space
of instantons. One of the main conceptual results of Ref.~\cite{local}
was the realization that it is actually unnecessary to integrate over
the whole moduli space of a constrained instanton, rather one can
expand around the exact solutions corresponding to point-like
instantons. In order to avoid the singular nature of point-like
instantons, Ref.~\cite{local} introduced a regularization based on the
smooth resolution of the instanton moduli space first described in
purely geometrical terms without reference to the gauge theory by
Nakajima in Ref.~\cite{Nakajima:1993jg}.\footnote{Subsequent to Nakajima, 
it was realized
by Nekrasov and Schwarz \cite{NS} that this smooth resolution of the
instanton moduli space arises naturally when the theory is defined on
a non-commutative spacetime.}
In this case the 
exact instanton solutions are smooth and of small, but
fixed, size. Since these exact solutions really do have a local action
we call them ``topicons''.

In order to uncover the ``calculus of topicons'', 
the key idea is to formulate the integral
over the original instanton moduli space as a zero-dimensional topological, or
cohomological, field theory.\footnote{Considering the context perhaps
``topological, or cohomological, matrix theory'' would be more
appropriate.} In this formalism, there exists a 
nilpotent fermionic symmetry, or---depending on taste---a BRST 
operator, and the integrals can be shown to localize on the critical
points of the potential; namely, the subspace of point-like
instantons. For $k$-instantons, this space is simply the space of $k$
indistinguishable points in ${\mathbb R}^4$, or ${\rm Sym}^k\,{\mathbb
R}^4$---a much simpler space than the full instanton moduli
space $\ms_{k,N}$. 

As alluded to above, 
the problem is that the subspace of point-like instantons arises as a
singular subspace of $\ms_{k,N}$ and has singularities of its own when
two, or more, points come together in ${\mathbb R}^4$. These
singularities complicate the issue of localization. In order to
resolve the difficulties, in Ref.~\cite{local}, we suggested that one
could consider the spacetime non-commutative version of the $\U(N)$
theory. After all, the instanton moduli space in the non-commutative
theory, $\ms_{k,N}^{(\zeta)}$, is a resolved version of $\ms_{k,N}$
which no longer has singularities \cite{NS}. 
Physically instantons can no longer shrink to zero size and the
consequences of Derrick's Theorem are avoided. In fact in the
non-commutative theory there are now exact non-singular instanton
solutions even in the presence of VEVs: the topicons. 
Actually there are $N$ flavours of
topicon obtained by embedding the spacetime non-commutative $\U(1)$
instanton solutions  
in each of $N$ unbroken abelian factors of the gauge group. (For a
discussion of instantons in the non-comutaive $\U(1)$ theory see
Refs.~\cite{NS,Furuuchi123,Nekrasov:2000zz,Schwarz:2001ru} and
references therein.)

So the following picture emerges. For instanton charge $k$, the exact
instanton solutions come as a disjoint union of spaces
associated to the inequivalent partitions $k\to k_1+\cdots+k_N$,
$k_u\geq0$, where each $k_u$ corresponds to each of $N$ $\U(1)$
subgroups picked out by the VEV. Hence, the space of exact solutions,
or moduli space of topicons, lying within the 
larger moduli space is of the form
\EQ{
\ms_{k,N}\ \overset{\text{resolve}}\longrightarrow\ 
\ms_{k,N}^{(\zeta)}\ \overset{\text{exact}}\supset\ 
\ms_{k,N}^{(\zeta)}\Big|_{\text{topicon}}=
\bigcup_{\text{partitions}\atop 
k_1+\cdots+k_N}\ms^{(\zeta)}_{k_1,1}\times
\cdots\times\ms^{(\zeta)}_{k_N,1}\ .
\label{partitions}
}
Each factor $\ms_{k,1}^{(\zeta)}$ is a smooth resolution of the
singular space $\ms_{k,1}={\rm Sym}^k\,{\mathbb R}^4$.
The picture that we have arrived at qualitatively is one of the
main predictions of the localization technique as described in the
companion paper \cite{local}. 

In this paper, following on from \cite{local}, 
we consider the problem of calculating the
prepotential in the theory obtained by softly breaking the $\N=4$
supersymmetric gauge theory to $\N=2$ by adding mass terms: the
so-called ``$\N=2^*$ theory''. Using localization we will be able to
completely evaluate both the one- and two-instanton contributions to the
prepotential of the $\SU(N)$ theory (or more precisely the $\U(N)$ 
spacetime non-commutative version) and the results
that we obtain will be consistent with the same quantities extracted
from the Seiberg-Witten curve. More
ambitiously we will calculate the leading order contribution to the
prepotential in an expansion in the supersymmetry breaking mass to all
orders in the instanton charge. Once again we find perfect agreement
with the predictions of Seiberg-Witten theory. 
This consistency relies on the hypothesis, first made in
Ref.~\cite{Hollowood:2001ng}, that introducing spacetime
non-commutativity as a device for regulating the singularities of the
instanton moduli space does not affect the instanton contributions to
the prepotential.\footnote{Roughly speaking point-like instantons do
not couple to the VEV and so resolving them does not lead to any
VEV-dependent corrections.
As in Ref.~\cite{local}, we shall find some physically
irrelevant, {\it i.e.\/}~VEV independent, differences between
the instanton contributions in the commutative and non-commutative
theories.}

We now describe the predictions of Seiberg-Witten theory. 
As usual, the prepotential of the $\N=2^*$ theory has the form
\EQ{
{\cal F}={\cal F}_{\text{pert}}
+\frac1{2\pi i}\sum_{k=1}^\infty{\cal F}_ke^{2\pi ik\tau}\ ,
\label{expf}
}
where the sum is over the contributions from instantons of charge
$k$. These contributions are simply given, up to a multiplicative
factor, by the 
{\it centred instanton partition function\/} \cite{MO-I,MO-II,local,MY}:
\EQ{
{\cal F}_k=-m^2\widehat{\EuScript Z}_{k,N}^{\sst(\N=2^*)}\ ,
\label{ndeq}
}
where $m$ is the supersymmetry breaking mass. The partition function
$\widehat{\EuScript Z}_{k,N}^{\sst(\N=2^*)}$ is
defined as an integral over the suitably supersymmetrized
version of the centred $k$-instanton moduli space
$\widehat\ms_{k,N}$.\footnote{The centred moduli space has the overall
position of the instanton configuration factored off:
$\ms_{k,N}={\mathbb R}^4\times\widehat\ms_{k,N}$.} If
$\Bomega^{\sst(\N=4)}$ is the $\N=4$ supersymmetric volume form, then
\EQ{
\widehat{\EuScript Z}_{k,N}^{\sst(\N=2^*)}=\int_{\widehat\ms_{k,N}}
\Bomega^{\sst(\N=4)}\,e^{-S-mS_{\text{mass}}}\ .
\label{coeff}
}
The quantity $S$ is the instanton effective action which 
depends on the VEVs parameterizing the Coulomb branch and we find it
convenient to explicitly separate out the terms $S_{\text{mass}}$ which
describe the effect of adding the supersymmetry breaking mass $m$. 
The fact that there is a non-trivial action on the instanton moduli
space is a direct manifestation of the 
fact that instantons are
not exact solutions of the equation-of-motion when the scalar fields
have VEVs: 
rather they should be treated as constrained instantons {\it \`a la\/}
Affleck \cite{Affleck,ADS}.

The first two instanton contributions were extracted from the 
Seiberg-Witten curve for this theory via a procedure making extensive use of 
the underlying Calogero-Moser
integrable system in Ref.~\cite{DP}. The predictions in our notation
read:\footnote{The following expressions are identical
to those in Eq.~(5.22) of Ref.~\cite{DP}. I am grateful to the authors of 
Ref.~\cite{DP} for pointing out a typo in their subsequent Eq.~(5.23).}
\AL{
{\cal F}_1&=-m^2\sum_{u=1}^N\prod_{v=1\atop(\neq
u)}^N\Big(1-\frac{m^2}{\phi_{uv}^2}\Big)\ ,\\
{\cal
F}_2&=-m^2\sum_{u=1}^N\Big(\tfrac32T_u(\phi_u)^2+\tfrac14
T_u(\phi_u)\frac{\partial^2
T_u(\phi_u)}{\partial\phi_u^2}\Big)\notag\\
&-m^4\sum_{u,v=1\atop 
(u\neq v)}^NT_u(\phi_u)T_v(\phi_v)\Big[\frac1{\phi_{uv}^2}-
\frac12\frac1{(\phi_{uv}+m)^2}-\frac12\frac1{(\phi_{uv}-m)^2}\Big]\ ,
\label{ktwo}
}
where we have defined $\phi_{uv}=\phi_u-\phi_v$ and make use of the
functions 
\EQ{
T_u(x)=\prod_{v=1\atop(\neq
u)}^N\Big(1-\frac{m^2}{(x-\phi_v)^2}\Big)\ .
\label{deft}
}

The instanton contributions for $k>2$ can---at least in principle---be
generated order-by-order in instanton number
from the recursion relation established in
\cite{Chan:2001nt}. However another approach, 
described in Ref.~\cite{MNW}, established
a very different kind of---and for us more useful---recursion relation
for the prepotential. In this work, the prepotential is expanded in
powers of $m^2$ rather than the instanton factor $e^{2\pi
i\tau}$. Up to irrelevant constant factors
\EQ{
{\cal F}=\frac1{2\pi i}\sum_{n=1}^\infty f_n(\tau,\phi_u)m^{2n}\ .
}
The recursion relation can then used to find $f_n$ in terms of $f_p$,
$p<n$. The first two terms of the expansion are
\AL{
f_1&=\tfrac12\sum_{u,v=1\atop(u\neq v)}^N\log\phi_{uv}^2\ ,\\
f_2&=-\frac{E_2(\tau)}{24}\sum_{u,v=1\atop(u\neq v)}^N\frac1{\phi_{uv}^2}\ ,
}
where $E_2(\tau)$ is the regulated Eisenstein series of weight
two. This function has an instanton expansion
\EQ{
E_2(\tau)=1-24\sum_{k=1}^\infty\Big(\sum_{d|k}d\Big)e^{2\pi
ik\tau}\ ,
}
where the sum is over the integer divisors $d$ of $k$.
Putting the mass expansion together with the expansion over instanton
number we find
\EQ{
{\cal F}_k=m^4\Big(\sum_{d|k}d\Big)\sum_{u,v=1\atop(u\neq
v)}^N\frac{1}{(\phi_u-\phi_v)^2}+{\cal O}(m^6)\ .
\label{impp}
}
Again we emphasize that this is modulo VEV-independent pieces.

\section{The $\N=4$ Instanton Calculus and Localization}

The primary goal of this section is to define the centred instanton
partition function and construct the fermionic BRST symmetry. We will
not provide a completely self-contained description of the  
instanton calculus: for this
one can refer to the more detailed treatments in 
Refs.\cite{KMS,MY,MO3} along with the companion paper \cite{local}.

The ADHM construction \cite{ADHM} involves a set of over-complete
collective coordinates that are subject to a set of non-linear
constraints. The variables consist of the $N\times k$ complex matrices
$w_\aD$, $\aD=1,2$, 
with elements $w_{ui\aD}$ and $k\times k$ traceless (in order to describe the
centred moduli space) Hermitian matrices $a'_n$.\footnote{$\SU(N)$
gauge indices are denoted $u,v=1,\ldots,N$ and ``instanton'' indices
are denoted $i,j=1,\ldots,k$, where $k$ is the instanton charge. In
addition $\bar w^\aD\equiv (w_\aD)^\dagger$.} Our conventions are
described in the Appendix. The ADHM constraints are
\EQ{
\tau^{c\aD}{}_\bD\big(\bar w^\bD w_\aD+\bar
a^{\prime\bD\alpha}a'_{\alpha\aD}\big)=\zeta^c1_{\sst[k]\times[k]}\ .
\label{adhm}
}
Here, $\zeta^c$ are real constants which parameterize the spacetime
non-commutativity of the gauge theory. For the commutative theory one
has $\zeta^c=0$. In order to regulate the singularities of the
instanton moduli space, we will consider for the most part the non-commutative
theory where $\zeta^c$ is non-trivial. 
Without-loss-of-generality, we will choose 
\EQ{
\zeta^1=\zeta^2=0\ ,\qquad\zeta^3\equiv\zeta>0\ .
\label{ncom}
}
To complete the construction of the instanton moduli space one takes a
quotient of the solutions of \eqref{adhm} by the $\U(k)$ symmetry that
acts as
\EQ{
w_\aD\to w_\aD\,{\cal U}\ ,\qquad a'_n\to{\cal U}^\dagger\,
a'_n\,{\cal U}\ ,
\label{symm}
}
${\cal U}\in\U(k)$. As described more fully in \cite{MY,Bruzzo:2001di},
the ADHM construction is an example of the
hyper-K\"ahler quotient construction \cite{Hitchin:1987ea} 
starting from flat space with metric
\EQ{
ds^2=
8\pi^2{\rm tr}_k\,\big[d\bar w^\aD\,dw_\aD+da'_n\,da'_n\big]\ .
\label{qum}
}
by the group action \eqref{symm}. 

In an $\N=4$ supersymmetric theory an instanton has a set of Grassmann
collective coordinates which parameterize the $8kN$ zero modes of the
Dirac operator in the instanton background.\footnote{For a review of
the $\N=4$ instanton calculus see Refs.~\cite{MO3,MY}.} In the supersymmetric
extension of the ADHM construction we first define a set of
over-complete Grassmann variables $\{\mu^A,\bar\mu^A,\CM^{\prime
A}_\alpha\}$. Here $A=1,\ldots,4$ is a fundamental index of the $\SU(4)$
$R$-symmetry. The quantity
$\mu^A$ is an $N\times k$ matrix with elements $\mu^A_{ui}$,
$\bar\mu^A$ is a $k\times N$ matrix with elements $\bar\mu^A_{iu}$ and
$\CM^{\prime A}_\alpha$ are traceless
$k\times k$ matrices with elements $(\CM^{\prime
A}_\alpha)_{ij}$. 
This over-complete set of variables is subject to fermionic
analogues of the ADHM constraints:
\EQ{
\bar\mu^A w_\aD+\bar w_\aD\mu^A+[\CM^{\prime\alpha A},a'_{\alpha\aD}]=0\ .
\label{fadhm}
}

As in the $\N=2$ theory with fundamental hypermultiplets
discussed in \cite{local}, it is convenient to introduce some
additional auxiliary variables. In the $\N=4$ context these consist of:
$\chi_a$, $a=1\ldots,6$, 
a six-vector\footnote{We shall often denote a six-vector by a
bold symbol, {\it e.g.\/}
$\Bchi$.} of Hermitian $k\times k$ matrices; $D^c$, $c=1,2,3$, 
three Hermitian $k\times k$ matrices; and $k\times k$ matrices
of Grassmann superpartners $\bar\psi^\aD_A$. Using these variables, the
$\N=4$ centred instanton partition function can be written in a 
completely ``linearized'' form:
\SP{
\widehat{\EuScript Z}_{k,N}^{\sst(\N=4)}
&=\frac{2^{-5/2k^2+k/2-2kN}\pi^{-10k^2-6kN+6}k^2}
{{\rm Vol}\,\U(k)}
\int d^{2kN}w\,d^{2kN}\bar w\,d^{4(k^2-1)}a'\,d^{3k^2}D\,
d^{6k^2}\Bchi\\
&\qquad\qquad\qquad\times\,d^{4kN}\mu\,d^{4kN}\bar\mu\,
d^{8(k^2-1)}\CM'\,d^{8k^2}\bar\psi\,\exp(-S)\ .
\label{ipf}
}
Here, the {\it instanton effective action\/} is\footnote{Our
conventions are set out in the Appendix.} 
\EQ{
S=4\pi^2{\rm tr}_k\Big\{\big|w_\aD\Bchi+\Bphi w_\aD\big|^2
-[\Bchi,a'_n]^2
+\tfrac 12\bar\BSigma_{AB}\cdot\big[\bar\mu^A(\mu^B\Bchi+\Bphi\,\mu^B)
+\CM^{\prime\alpha
A}\CM^{\prime B}_{\alpha}\Bchi\big]\Big\}+ 
S_{\text{L.m.}}\ .
\label{jslmm}
}
The variables  $D^c$
and $\bar\psi^\aD_A$ act as Lagrange multipliers
for the bosonic and fermionic ADHM constraints \eqref{adhm} and
\eqref{fadhm} through the final term in the action:
\EQ{
S_{\text{L.m.}}=-4i\pi^2{\rm
tr}_k\Big\{\bar\psi_A^\aD\big(\bar\mu^A 
w_\aD+\bar w_\aD\mu^A+[\CM^{\prime\alpha A},
a'_{\alpha\aD}]\big)+D^c\big(\tau^{c\aD}{}_\bD(\bar w^\bD w_\aD+\bar
a^{\prime\bD\alpha}a'_{\alpha\aD})-\zeta^c\big)\Big\}\ .
}
In \eqref{jslmm}, $\Bphi$ is the VEV of the scalar field. In the
$\N=4$ theory, before the mass perturbation is considered, $\Bphi$ is a
six-vector of diagonal (traceless) $N\times N$-dimensional matrices
with elements $\Bphi_u$, $u=1,\ldots,N$.

An expression for the $\N=4$ supersymmetric volume form on the
resolved centred instanton moduli space $\widehat\ms_{k,N}^{(\zeta)}$
is obtained by integrating out the the auxiliary variables
$\{\chi_a,D^c,\bar\psi^\aD_A\}$. Notice that after this has been done
the volume integral is weighted by an exponential of a non-trivial
function depending on the VEVs $\Bphi$. If
$\Bphi=0$ then this function does not vanish:
there remains a non-trivial quadrilinear
coupling of the Grassmann collective coordinates which was first
obtained for the $\SU(N)$ theory in Ref.\cite{MO3}. 
This is the major difference between
instantons in the $\N=4$ versus $\N=2$ theories: in the former,
instantons with Grassmann fields turned on are not generally exact
solutions of the equations-of-motion \cite{MY,MO3} in contrast to the
latter. This means that in the $\N=2$ theory with zero VEVs 
the centred instanton partition function vanishes
since there are unsaturated Grassmann integrals. In the $\N=4$ theory
with vanishing VEVs the centred instanton partition function does not vanish.
Actually it is precisely the 
Gauss-Bonnet-Chern integral on $\widehat\ms_{k,N}^{(\zeta)}$. The
reason is that the quadratic
coupling of the Grassmann collective coordinates involves the Riemann
tensor and performing the Grassmann integrals pulls down powers of the
Riemann tensor contacted in precisely the right way to give the 
Euler form. On a compact space, therefore, the partition function
(with zero VEVs)
would yield the Euler characteristic. However,
$\widehat\ms_{k,N}^{(\zeta)}$ is not compact since individual
instantons can become arbitrarily separated in ${\mathbb R}^4$. To
define the Euler characteristic on such a space, we have to cut it off;
for instance, on a large sphere of radius $R$:
\EQ{
{\tr }_k(\bar w^\aD w_\aD+a'_na'_n)=R^2\ .
} 
As $R\to\infty$, the 
Euler characteristic receives a bulk contribution given by the
Gauss-Bonnet-Chern integral and also a boundary contribution
\cite{EGH}. The Gauss-Bonnet-Chern integral for
$\widehat\ms_{1,N}^{(\zeta)}$ was
calculated explicitly in \cite{Dorey:2001zq} with result
\EQ{
\widehat\Z_{1,N}^{\sst(\N=4)}=
\frac{2\Gamma(N+\tfrac12)}{\Gamma(N)\Gamma(\tfrac12)}\
.
}
The Euler characteristic of $\widehat\ms_{1,N}^{(\zeta)}$ (in the
sense define above) has been computed by Nakajima using Morse theory
\cite{Nakajima:1993jg,Nakajima:1996ka} giving the result $N$. This
means that the boundary contribution on the large sphere at infinity must be 
non-vanishing.

When the VEVs are turned on, the instanton effective action \eqref{jslmm}
is modified: instantons
become constrained and there is a non-trivial potential on the moduli
space. On the mathematical side the resulting situation is rather
familiar in the context of the Mathai-Quillen formalism for dealing
with representations of Euler classes of vector bundles \cite{matquil}
(for a physicists' review see Ref.~\cite{Blau:1995rs}.) The VEVs
correspond to a vector fields on $\widehat\ms_{k,N}^{(\zeta)}$ and the
resulting integral over the moduli space involves the Mathai-Quillen
form. In the compact case this form is cohomologous to the Euler class
and so yields the Euler characteristic. By making the VEVs very large
the integral localizes on the critical points of the vector
fields. Using this formalism it is possible to show that the Euler
characteristic can be computed either as a bulk integral (the
Gauss-Bonnet Theorem) or as a sum over the critical points (the
Poincar\'e-Hopf Theorem). In the following we will basically set up
this formalism using the language of topological, or cohomological
field theory, where one uses Grassmann variables and BRST operators
rather than forms and exterior derivatives. However, one should keep
in mind that our application is rather more complicated than usual
due to non-compactness.

Before we proceed, it is useful to take stock of the various
symmetries that play a r\^ole in the $\N=4$ instanton
calculus. Firstly, we have the $\SO(4)$ Euclidean Lorentz group with
covering group $\SU(2)_L\times\SU(2)_R$. The $\alpha=1,2$ ($\aD=1,2$) 
indices are spinor indices of $\SU(2)_L$ ($\SU(2)_R$). Then we have the
$\SU(4)\simeq\SO(6)$, $R$-symmetry group with spinor indices
$A=1,2,3,4$ and vector indices $a=1,\ldots,6$. Global $\U(N)$ gauge
transformations act on quantities with indices $u=1,\ldots,N$ and
finally the ADHM construction involves the auxiliary $\U(k)$ symmetry
which acts on indices $i=1,\ldots,k$.

To describe the soft-breaking to $\N=2$ by mass terms, we first have
to restrict the non-zero elements of the $\SO(6)$ vector of VEVs, $\phi_a$, to
$a=1,2$. Since the mass terms break $\N=4$ to $\N=2$, the $R$-symmetry
is broken from $\SU(4)$ to $\SU(2)\times\U(1)$. In order to describe
the symmetry breaking, it
is expedient to decompose $\SO(6)$ vector indices as
\EQ{
a\to (a',\hat a)\ ,\qquad a'=1,2,\qquad\hat a=3,4,5,6\ ,
}
and $\SU(4)$ spinor indices as\footnote{Where necessary we will raise and lower
$A'$ and $\hat A$ indices with the $\epsilon$ tensor with 
$\epsilon_{21}=\epsilon^{12}=\epsilon_{43}=\epsilon^{34}=1$, 
$\epsilon_{12}=\epsilon^{21}=\epsilon_{34}=\epsilon^{43}=-1$.} 

\EQ{
A\to (A',\hat A)\ ,\quad A'=1,2,\qquad \hat A=3,4\ .
}
The unbroken $\SU(2)$ factor of the $R$-symmetry is then indexed by $A'$.
The mass deformation is obtained by adding 
the following term to the instanton effective action \cite{DKMn4,MY}:
\EQ{
mS_{\text{mass}}=-\pi^2M_{AB}{\rm
tr}_k\big(2\bar\mu^A\mu^B+\CM^{\prime\alpha A}\CM^{\prime
B}_{\alpha}\big)\ ,
}
where
\EQ{
M_{AB}=m\MAT{0&\a0&\a0&\a0\\ 0 &\a0&\a0&\a0 \\ 0&\a0&\phantom{-}1&\a0\\
0&\a0 &\a0& \phantom{-}1}\ .
}
In order that the mass deformation preserves $\N=2$ supersymmetry, the
components $\phi_{\hat a}$ of the VEV must vanish. 
The remaining variables $\phi_{a'}$, $a'=1,2$, parameterize the
Coulomb branch of the $\N=2^*$ theory. We will find that the
prepotential depends holomorphically on the variables defined by
\EQ{
\phi=i\phi_1-\phi_2\equiv-\tfrac i2\epsilon^{\hat A\hat B}\bar\BSigma_{\hat A\hat B}\cdot\Bphi\ .
\label{defphi}
}
Here, $\phi$ is then a diagonal $N\times N$ traceless matrix with
elements $\phi_u$.With the mass term $S_{\text{mass}}$ 
added to the instanton effective action we have
\EQ{
\widehat\Z_{k,N}^{\sst(\N=4)}\longrightarrow
\widehat\Z_{k,N}^{\sst(\N=2^*)}=\int_{\widehat\ms_{k,N}}
\Bomega^{\sst(\N=4)}\,e^{-S-mS_{\text{mass}}}
}
which gives the $k$-instanton coefficient of the prepotential via \eqref{ndeq}.

Before mass deforming, the instanton effective action \eqref{jslmm} 
is invariant under eight supersymmetries corresponding to 
precisely half the number of the $\N=4$ theory reflecting the fact
that an instanton is a BPS configuration that breaks half the
supersymmetries of the underlying field theory.
On the full set of variables, the transformations are
\ALT{2}{
\delta a'_{\alpha\aD}&=  i\bar{\xi}_{\dot{\alpha}A}
{\cal M}^{\prime A}_{\alpha}\ ,\qquad& \delta{\cal M}^{\prime
A}_{\alpha}&=  -2i\bar\xi^{\aD}_B \BSigma^{AB}\cdot
[a'_{\alpha\aD},\Bchi]\ ,\notag\\
\delta w_\aD&= i\bar\xi_{\aD A} \mu^A\ ,\qquad&
\delta \mu^A&=  -2i\bar\xi^{\aD}_B\BSigma^{AB}\cdot
\big(w_\aD\Bchi+\Bphi w_\aD\big)\ ,\notag\\
\delta\Bchi&=-\BSigma^{AB}\bar\xi_{\aD A}\bar\psi^\aD_B\ ,\qquad&
\delta \bar\psi_{A}^\aD& =
2\bar\Sigma_{abA}{}^B
[\chi_a,\chi_b]\bar\xi_{B}^\aD-iD^c\tau^{c\aD}{}_\bD 
\bar\xi_{A}^\bD\ ,\label{susy22}\\
\delta D^c&= -i\tau^{c\aD}{}_\bD\bar\xi_{\aD
 B}\BSigma^{AB}\cdot[\bar\psi^\bD_A,\Bchi]\ .&&
\notag
}
Once we have added the mass term, and set $\phi_{\hat a}=0$, 
only four supersymmetries corresponding to $\bar\xi_{\aD A'}$ remain
as symmetries.

One can interpret the partition function $\widehat\Z_{k,N}^{\sst(\N=4)}$ 
as the dimensional reduction of an
$\N=(0,1)$ supersymmetric gauged linear $\sigma$-model in six
dimensions \cite{MO3,MY}. In this
interpretation $\Bchi$ is the $\U(k)$ six-dimensional gauge field
forming a vector multiplet of supersymmetry along with
$\bar\psi_A^\aD$ and $D^c$ and the non-commutativity parameters
$\zeta^c$ arise as Fayet-Illiopolos terms for the abelian subgroup of $\U(k)$. 
These variables have no kinetic term (in
six dimensions) and on integrating them out 
one recovers a non-linear $\sigma$-model with the hyper-K\"ahler space 
$\widehat\ms_{k,N}^{(\zeta)}$ as target.

The fermionic symmetry, or ``BRST operator'', for the $\N=4$ instanton
calculus was constructed in \cite{Dorey:2001zq}.\footnote{For related work in
the context of the $\N=2$ with fundamental hypermultiplets see
Refs.~\cite{local,Fucito:2001ha}.} It is simply given by
a certain combination of the supersymmetries.
Firstly, from the supersymmetry transformations we can define corresponding 
supercharges via
$\delta=\xi_{\aD A}Q^{\aD A}$. The fermionic symmetry we are after is
then generated by the BRST operator 
\EQ{
\Q=\epsilon_{\aD A'}Q^{\aD A'}
}
which gives
\ALT{2}{
\Q\, a'_{\alpha\aD}&=  i\epsilon_{\dot{\alpha}A'}
{\cal M}^{\prime A'}_{\alpha}\ ,\qquad&
 \Q\,{\cal M}^{\prime A}_{\alpha} &=  -2i\delta\xi^{\aD}{}_{B'}
\BSigma^{AB'}\cdot
[a'_{\alpha\aD},\Bchi]\ ,\notag\\
\Q\, w_\aD&= i\epsilon_{\aD A'} \mu^{A'}\ ,\qquad&\Q\, \mu^A&=  -2i\delta^\aD{}_{B'}\BSigma^{AB'}\cdot
\big(w_\aD\Bchi+\Bphi w_\aD\big)\ ,\notag\\
\Q\,\Bchi&=-\BSigma^{A'B}\epsilon_{\aD A'}\bar\psi^\aD_B\ ,\qquad&
\Q\, \bar\psi_{A}^\aD &=
2\bar\Sigma_{abA}{}^{B'}
[\chi_a,\chi_b]\delta^\aD{}_{B'}-iD^c\tau^{c\aD}{}_\bD 
\delta^\dD{}_{A'}\ ,\label{qact}\\
\Q\, D^c&= -i\tau^{c\aD}{}_\bD\epsilon_{\aD
 B'}\BSigma^{AB'}\cdot[\bar\psi^\bD_A,\Bchi]\ .
\notag
}
Notice that the definition of $\Q$ mixes up spacetime and $R$-symmetry
indices as is characteristic of topological twisting. 
It can be shown that $\Q$ is nilpotent up to an infinitesimal 
$\U(k)\times\SU(N)$ transformation generated by $\chi$ and $\phi$:
\EQ{
\Q^2\,\ast=2i\big(\delta_\chi\,\ast\,+\,\delta_\phi\,\ast\big)\ ,
}
where $\delta_\chi$ and $\delta_\phi$ are infinitesimal 
$\U(k)$ and $\U(N)$ transformations generated by  
\EQ{
\chi=i\chi_1-\chi_2\equiv -\tfrac i2\epsilon^{\hat A\hat B}
\bar\BSigma_{\hat A\hat B}\cdot\Bchi\ ,
\label{moom}
}
and $\phi$ in \eqref{defphi},
respectively, {\it
e.g.\/}~$\delta_\chi\,w_\aD=w_\aD\chi$ and $\delta_\phi\,w_\aD=\Bphi\,
w_\aD$. 

It is now possible to write the instanton effective action \eqref{jslmm} in a
manifestly $\Q$-exact way. This was done in
Ref.~\cite{Dorey:2001zq} in a slightly more general form where
the variables $\{\chi_a,D^c,\bar\psi_A\}$ have ``kinetic terms'' with
a coupling constant $g_0^{-2}$
rather than being auxiliary. However, the kinetic terms can be removed
by a careful re-scaling by $g_0$,
and then taking the limit $g_0\to\infty$. We now describe the result
of this procedure. The construction is greatly facilitated by
introducing some additional auxiliary variables $\{H_\alpha^{\hat A},F^{\hat A},\bar
F^{\hat A}\}$ which linearize the fermionic symmetry:
\SP{
\Q\,\CM^{\prime \hat A}_\alpha&=H^{\hat A}_\alpha\ ,\qquad\Q\,
H^{\hat A}_\alpha=2i[\CM^{\prime \hat A}_\alpha,\chi]\ ,\\
\Q\,\mu^{\hat A}&=F^{\hat A}\ ,\qquad\Q\,
F^{\hat A}=2i(\mu^{\hat A}\chi+\phi\mu^{\hat A})\ ,\\
\Q\,\bar\mu^{\hat A}&=\bar F^{\hat A}\ ,\qquad\Q\,
\bar F^{\hat A}=-2i(\chi\bar\mu^{\hat A}+\bar\mu^{\hat A}\phi)\ .
\label{new}
}
One can then show that the instanton effective action is $\Q$-exact:
\EQ{
S=\Q\,\Xi\ ,
}
where
\SP{
\Xi&=4\pi^2{\rm tr}_k\Big\{\tfrac12\delta^{A'}{}_{\aD}
\bar w^\aD\bar\BSigma_{A'B'}\cdot\big(\mu^{B'}\Bchi
+\Bphi\,\mu^{B'}\big)+\tfrac14\delta^{A'}{}_{\aD}\bar
a^{\prime\aD\alpha}\bar\BSigma_{A'B'}\cdot[\CM^{\prime
B'}_\alpha,\Bchi]\\
&\qquad+\tfrac14\CM^{\prime \hat A}_\alpha\big(\tfrac12 
H^\alpha_{\hat A}-h_{\hat A}^\alpha\big)
+\tfrac14\bar\mu^{\hat A}\big(\tfrac12F_{\hat A}-f_{\hat A}\big)+
\tfrac14\big(\tfrac12\bar F^{\hat A}-\bar f^{\hat A}\big)\mu_{\hat A}\\
&\qquad\qquad+\delta^{A'}{}_\aD\bar\psi^\bD_{A'}\big(\bar w^\aD
w_\bD+\bar
a^{\prime\aD\alpha}a'_{\alpha\bD}-\tfrac12\tau^{c\aD}{}_\bD\zeta^c\big)\Big\}\ .
}
In the above, 
we have defined the following quantities (the 
``equations'' in the language of cohomological field theory)
\SP{
h^{\hat A}_\alpha&=-2i\delta^\aD{}_{A'}\BSigma^{\hat AA'}\cdot[a'_{\alpha\aD},\Bchi]\ ,\\
f^{\hat A}&=-2i\delta^\aD{}_{A'}\BSigma^{\hat AA'}\cdot\big(w_\aD\Bchi+\Bphi\,w_\aD\big)\ ,\\
\bar f^{\hat A}&=-2i\epsilon_{\aD A'}
\BSigma^{\hat AA'}\cdot\big(\Bchi
\bar w^\aD+\bar w^\aD\Bphi\big)\ .
}
More precisely, to obtain \eqref{jslmm}, one has to integrate out the 
variables
$\{H^{\hat A}_\alpha,F^{\hat A},\bar F^{\hat A}\}$ which appear
quadratic in the
action. This is equivalent to setting 
\EQ{
H^{\hat A}_\alpha\to h^{\hat A}_\alpha\ ,\qquad F^{\hat A}\to f^{\hat A}\ ,\qquad
\bar F^{\hat A}\to \bar f^{\hat A}\ .
}

We have, up till now, not discussed the $\N=4\to2$ breaking mass term
$S_{\text{mass}}$. The question is how this term modifies the picture we
have established of the BRST operator $\Q$ and a $\Q$-exact action? The answer
involves a mass-dependent deformation of the BRST operator itself which we
denote $\Q_m$, while $\Xi$ remains unaffected:
\EQ{
S+mS_{\text{mass}}=Q_m\Xi\ .
}
For this to work we must
set $\phi_{\hat a}=0$ which ensures that the resulting set-up has
$\N=2$ supersymmetry. The action of the deformed symmetry is equal to
\eqref{qact} and \eqref{new} up to the following changes:
\SP{
 \Q_m\,H^{\hat A}_\alpha&=2i[\CM^{\prime \hat A}_\alpha,\chi]+
\varpi^{\hat A}{}_{\hat B}\CM^{\prime\hat B}_\alpha\\
\Q_m\,F^{\hat A}&=2i(\mu^{\hat A}\chi+\phi\mu^{\hat A})+\varpi^{\hat A}{}_{\hat B}\mu^{\hat B}\ ,\\
\Q_m\,\bar
F^{\hat A}&=-2i(\chi\bar\mu^{\hat A}+\bar\mu^{\hat A}\phi)+\varpi^{\hat A}{}_{\hat B}\bar\mu^{\hat B}\ ,\\
 \Q_m\, \bar\psi_{\hat A}^\aD &=
2\bar\Sigma_{ab\hat A}{}^{B'}
[\chi_a,\chi_b]\delta^\aD{}_{B'}+\bar\Sigma_{a\hat AB'}\epsilon^{\aD B'}
\varpi_a{}^b\chi_b\ ,
\label{neww}
}
where $\varpi$ is a specific, mass-dependent, infinitesimal generator
of a transformation in the
unbroken $\SU(2)\times\U(1)$ $R$-symmetry group. For the spinor and
vector representations of $\SU(4)$, respectively,
\EQ{
\varpi^A{}_B=\frac m8\MAT{0&\a0&\a0&\a0\\ 
0&\a0&\a0&\a0\\ 0&\a0&\a0&-1\\
0&\a0&\phantom{-}1&\a0}\ ,\qquad
\varpi_a{}^b=\frac m8
\MAT{0&\a0&\a0&\a0&\a0&\a0\\ 0&\a0&\a0&\a0&\a0&\a0\\ 0&\a0&\a0&\a0&-1&\a0\\
0&\a0&\a0&\a0&\a0&\phantom{-}1\\ 0&\a0&\phantom{-}1&\a0&\a0&\a0\\
0&\a0&\a0&-1&\a0&\a0}\ .
}
Note that $\varpi_a{}^b\phi_b=0$, as it should so that the mass
deformation is consistent with the VEV.

Now one can show that $\Q_m$ is nilpotent up to infinitesimal transformations
in the $\U(k)$ and $\U(N)$ symmetry groups, as before, but now, in
addition, an infinitesimal transformation in the unbroken 
$\SU(2)\times\U(1)$ $R$-symmetry group with generator $\varpi$. 
It is a standard argument to show that the
partition function---at least formally---localizes on the critical
points of the action $\Q_m\Xi$. Consider the more 
general integral
\EQ{
\widehat{\EuScript Z}_{k,N}^{\sst(\N=2^*)}(s)
=\int_{\widehat\ms_{k,N}}\Bomega^{\sst(\N=4)}\,\exp\big(
-s^{-1}\Q_m\,\Xi\big)\ .
\label{suggvg}
}
We then have
\EQ{
\PD{\widehat{\EuScript Z}_{k,N}^{\sst(\N=2^*)}(s)
}{s}=
s^{-2}\int_{\widehat{\ms_k}}\Bomega^{\sst(\N=4)}\,\Q_m
\Big\{\Xi\,\exp\big(-s^{-1}\Q_m\,\Xi\big)\Big\}\ ,
\label{varet}
}
using the fact that $\Q_m^2\,\Xi=0$.  
Since the volume form is invariant under supersymmetry (as proved in
Ref.~\cite{MY,DHKM}), $\U(N)$, $\U(k)$ and $\SU(4)$ $R$-symmetry, it is
$\Q_m$-invariant and so the
right-hand side of \eqref{varet} 
vanishes. Consequently, $\widehat{\EuScript Z}_{k,N}^{\sst(\N=2^*)}(s)$
is independent of $s$ and, therefore, it can be evaluated
in the limit $s\to0$ where the integral is dominated by
the critical points
of $\Q_m\,\Xi$. Since the result is independent of $s$,
under favourable circumstances---which will be shown to hold
in the present application---the Gaussian approximation is exact 
(for references to this kind of localization in the physics literature see
Refs.~\cite{Blau:1995rs,berline,Niemi:1994ej,Schwarz:1997dg} and references 
therein).

Notice in this formalism the dependence on the
anti-holomorphic component of the VEV $\phi^\dagger$ and the
non-commutativity couplings $\zeta^c$, resides solely in $\Xi$. Hence
the derivative of the integral with respect to either of these
parameters is $\Q_m$-exact and so the integral cannot depend on either
$\phi^\dagger$ and $\zeta^c$. On the other hand, the dependence of the
integrals on the holomorphic component of the VEVs $\phi$ and the mass
$m$ is through the operator $\Q_m$ and so there is every reason to
expect the integrals to depend on these parameters, as we will find.

Having established the idea of localization, we now
investigate exactly on what submanifolds the integrals localize.
The critical points are the zeros of
\eqref{jslmm},
\EQ{
\big|w_\aD\Bchi+\Bphi w_\aD\big|^2
-[\Bchi,a'_n]^2\ ,
}
which requires
\EQ{
w_\aD\Bchi+\Bphi w_\aD=[\Bchi,a'_n]=0\ .
}
Up to the $\U(k)$ auxiliary symmetry, there are a set of discrete
critical-point sets associated to the inequivalent partitions
\EQ{
k\to k_1+k_2+\cdots+k_N\ .
}
For a given partition, each $i\in\{1,2,\ldots,k\}$ is 
associated to a given $u$ by a map $u_i$ as follows:
\SP{
&\Big\{\underbrace{1,2,\ldots,k_1}_{u=1},
\underbrace{k_1+1,\ldots,k_1+k_2}_{u=2},\ldots,\\
&\qquad\ldots,\underbrace{k_1+\cdots+k_{u-1}+1,
\ldots,k_1+\cdots+k_u}_u,\ldots,\ldots,
\underbrace{k_1+\cdots+k_{N-1}+1,\ldots,k}_{u=N}\Big\}
}
and the variables have a block diagonal-form
\EQ{
\Bchi_{ij}=-\Bphi_{u_i}\delta_{ij}\ ,\qquad
w_{ui\aD}\propto \delta_{uu_i}\ ,\qquad
(a'_n)_{ij}\propto\delta_{u_iu_j}\ .
\label{critp}
}
The critical-point sets have a very suggestive form. Imposing the
ADHM constraints implies that in the $u^{\rm th}$ 
block the constraints are those of $k_u$ instantons in a
non-commutative $\U(1)$ gauge theory.
The critical submanifold associated to $\{k_1,\ldots,k_N\}$ is then simply
\EQ{
\frac{\ms^{(\zeta)}_{k_1,1}\times
\cdots\times\ms^{(\zeta)}_{k_N,1}}{{\mathbb R}^4}\ ,
\label{gfps}
}
where the quotient is by the overall centre of the
instanton. The factors $\ms^{(\zeta)}_{k,1}$ are the $k$-instanton moduli
space in the spacetime non-commutative theory with gauge group
$\U(1)$. As we mentioned in the introduction these spaces are smooth
resolutions of the symmetric product ${\rm Sym}^k\,{\mathbb R}^4$.
We interpret \eqref{gfps} as the moduli space of a composite configuration of 
topicons involving $k_u$ of flavour $u$.

\section{One Instanton}

We now use the localization technique to evaluate the centred one-instanton
partition function. The details are very similar to the $\N=2$ case
with fundamental hypermultiplets described in Ref.~\cite{local}.

The instanton effective action has $N$
critical points, corresponding to a single topicon of arbitrary flavour
labelled by $v\in\{1,2,\ldots,N\}$, at which \eqref{critp}
\EQ{
\Bchi=-\Bphi_v\ ,\qquad w_{u\aD}\propto\delta_{uv}\ .
\label{critpn}
}
Note that $a'_n=0$ in the one-instanton sector.
With the choice of non-commutativity parameters in \eqref{ncom}, 
the ADHM constraints
\eqref{adhm} are solved with
\EQ{
w_{u\aD}=\sqrt{\zeta}e^{i\theta}\delta_{uv}\delta_{\aD1}\ ,
}
for an arbitrary phase angle $\theta$. The integrals over $w_{v\aD}$ are
then partially saturated by the $\delta$-function
ADHM constraints that arise once $D^c$
are integrated out. A trivial integral 
over the phase angle $\theta$ remains:
\EQ{
\int d^2w_v\,d^2\bar
w_v\,\prod_{c=1}^3\delta\big(\tfrac12\tau^{c\aD}{}_\bD(
\bar w_v^\bD w_{v\aD}-\zeta\delta^{c3})\big)
=8\pi\zeta^{-1}\ .
\label{yuppa}
}
Once the Lagrange multipliers $\bar\psi^\aD_A$ are integrated out the
resulting Grassmann $\delta$-functions saturate the 
the integrals over $\{\mu^A_v,\bar\mu^A_v\}$:
\EQ{
\int\,d\mu^A_v\,d\bar\mu^A_v\,\prod_{\aD=1}^2\delta\big(
\bar w_{v\aD}\mu^A_v+w_{v\aD}\bar\mu^A_v\big)=\zeta\ ,
\label{yuppb}
}
for each $A=1,\ldots,4$.
The remaining variables, $\{w_{u\aD},\mu^A_{u},\bar\mu^A_u\}$, $u\neq
v$, are all treated as Gaussian fluctuations around the critical point.
To this order, the instanton effective action \eqref{jslmm} is
\EQ{
S=4\pi^2\bigg\{\zeta\Bchi^2+\sum_{u=1\atop(\neq v)}^N
\Big(|\phi_{uv}|^2\big|w_{u\aD}\big|^2
+\tfrac 1{2}\bar\mu^A_u\big(\bar\BSigma\cdot\Bphi_{uv}-M\big)_{AB}
\mu^B_{u}\Big)\bigg\}+\cdots\ .
}
where $\Bphi_{uv}\equiv\Bphi_u-\Bphi_v$.
The integrals are easily done. Note that the integral over $\Bchi$
yields a factor of $\zeta^{-3}$ which cancels against the factors of
$\zeta$ arising from \eqref{yuppa} and \eqref{yuppb} so the final
result is, as expected, independent of $\zeta$. Summing over the $N$ 
critical-point sets 
gives the centred one-instanton partition function
\EQ{
\widehat{\EuScript Z}_{1,N}^{\sst(\N=2^*)}=\sum_{v=1}^N\prod_{u=1\atop(\neq
v)}^N\frac{{\rm det}_4(\bar\BSigma\cdot\Bphi_{uv}-M)}{\Bphi^4_{uv}}=
\sum_{v=1}^N\prod_{u=1\atop(\neq
v)}^N\Big(1-\frac{m^2}{(\phi_v-\phi_u)^2}\Big)\ .
\label{oneres}
}
Notice that the resulting expression is holomorphic in 
$\phi$ and independent of $\zeta^c$ as expected. 

When $m=0$, our result \eqref{oneres}, is simply the integer $N$:
precisely the Euler
characteristic of $\widehat\ms_{1,N}^{(\zeta)}$. This is because the
VEV is equivalent to introducing a Morse potential on the moduli
space in the language of the Mathai-Quillen formalism. 
In the one-instanton example the critical-point set is a set of
discrete points, precisely $N$ of them, and therefore the centred 
instanton partition function computes the Euler characteristic. For
$k>1$ the critical-point set will include non-compact components and
so the partition function will no longer compute a topological index.

\section{Two Instantons}

We now evaluate the centred two-instanton partition function
using localization. Once again some of the details are similar to the
$\N=2$ with fundamental hypermultiplets discussed in \cite{local}.

There are two kinds of critical submanifolds. The first in which
$u_1<u_2$ and the second when $u_1=u_2$; {\it i.e.\/}~two
topicons of different flavours and two of the same flavour,
respectively. We consider the contributions
in the next two subsections.

\subsection{Topicons of different flavour}

For two topicons of flavour $u_1$ and $u_2$, with $u_1<u_2$, 
the critical submanifold is
\EQ{
\ms_{1,1}^{(\zeta)}\times\ms_{1,1}^{(\zeta)}/{\mathbb R}^4\ .
}
On this submanifold, the 
ADHM constraints are solved with
\EQ{
w_{ui\aD}=\sqrt\zeta e^{i\theta_i}\delta_{uu_i}\delta_{\aD1}
\ ,\qquad
a'_n=\tfrac12\MAT{Y_n&0\\ 0&-Y_n}\ .
\label{diagsol}
}
The two phase angles $\theta_i$, $i=1,2$, are not genuine moduli since they can
be separately rotated by transformations in the subgroup
$\U(1)^2\subset\U(2)$ of the 
auxiliary group. The variables $Y_n$ are the genuine moduli
representing the relative positions of the two topicons.
The corresponding solution of the fermionic ADHM constraints 
\eqref{fadhm} on the critical submanifold is
\EQ{
\mu^A=\bar\mu^A=0\ ,\qquad\CM^{\prime A}_\alpha=\tfrac12\MAT{\xi^A_\alpha&0
\\ 0&-\xi^A_\alpha}\ ,
\label{fdiagsol}
}
where $\xi^A_\alpha$ are the eight relative supersymmetric modes of
the two topicons.

Including the fluctuations around the critical-point solution, we write
\EQ{
a'_n=\MAT{\tfrac12Y_n&[a'_n]_{12}\\ [a'_n]_{21}&-\tfrac12Y_n}\ ,
\qquad\CM^{\prime A}_\alpha=
\MAT{\tfrac12\xi^A_\alpha&[\CM^{\prime A}_\alpha]_{12}
\\ [\CM^{\prime A}_\alpha]_{21}&-\tfrac12\xi^A_\alpha}\ .
}
In addition, we have the following fluctuations 
$[w_\aD]_{pq}\equiv w_{u_pq\aD}$,
$[\mu^A]_{pq}\equiv \mu^A_{u_pq}$
and $[\bar\mu^A]_{pq}\equiv\bar\mu^A_{p u_q}$, for
$p,q=1,2$ and $2,1$, as well as the auxiliary variables $\Bchi$. 
It is convenient to make the shift
\EQ{
\Bchi\to\Bchi
-\MAT{\Bphi_{u_1}&0\\ 0&\Bphi_{u_2}}\
,\qquad\Bchi=\MAT{[\Bchi]_1&[\Bchi]_{12}\\ [\Bchi]_{21} & [\Bchi]_2} 
}
so that $\Bchi=0$ on the critical submanifold.
We then integrate over the Lagrange multipliers $D^c$ and
$\bar\psi_A^\aD$ which impose the ADHM constraints \eqref{adhm} and
\eqref{fadhm}. The two diagonal components of the constraints (in $i,j$
``instanton'' indices) are the ADHM constraints of the two single topicons.
The off-diagonal components vanish on the critical-point set and must
therefore be expanded to linear order in the fluctuations. Before
we write down the constraints, it is necessary to weed-out 
the fluctuations which correspond to $\U(2)$ ``gauge transformations''
of the critical-point solution. This can done
by imposing the condition that the fluctuations are orthogonal to 
infinitesimal $\U(2)$ transformations acting on the critical-point
solution which lie in the coset $\U(2)/\U(1)^2$. 
In turn this is done by inserting the following 
$\delta$-functions and Jacobian into the partition function:
\EQ{
{\rm Vol}\Big[\frac{\U(2)}{\U(1)^2}\Big]\big(\zeta+Y^2\big)^2\cdot
\prod_{p,q=1\atop(p\neq q)}^2\delta\big(\sqrt\zeta e^{i\theta_q}
[\bar w^1]_{pq}+\sqrt\zeta e^{-i\theta_p}[w^2]_{pq}+(-1)^p[\bar
a^{\prime\aD\alpha}]_{pq}Y_{\alpha\aD}\Big)\ .
\label{gfj}
}
When these gauge-fixing conditions are put together with the ADHM
constraints they can be written in a unified way:
\AL{
\sqrt\zeta e^{i\theta_q}[\bar w^\aD]_{pq}
+(-1)^p[\bar a^{\prime\aD\alpha}]_{pq}Y_{\alpha 1}&=0\ ,\label{bo1}\\
\sqrt\zeta e^{-i\theta_p}[w^\aD]_{pq}+(-1)^p[\bar a^{\prime\aD\alpha}]_{pq}
Y_{\alpha2}&=0\ ,\label{bo2}
}
for $p,q=1,2$ and $2,1$.
In the Grassmann sector, the off-diagonal fermionic ADHM constraints are
\AL{
\sqrt\zeta e^{i\theta_q}[\bar\mu^A]_{pq}+(-1)^p
[\CM^{\prime\alpha
A}]_{pq}Y_{\alpha1}&=(-1)^p[a'_{\alpha1}]_{pq}\xi^{\alpha A}\
,\label{oo1}\\ 
\sqrt\zeta e^{-i\theta_p}[\mu^A]_{pq}+(-1)^p[\CM^{\prime\alpha A}]_{pq}
Y_{\alpha2}&=(-1)^p[a'_{\alpha2}]_{pq}\xi^{\alpha A}\ ,\label{oo2}
}
where $Y_{\alpha\aD}=Y_n\sigma_{n\alpha\aD}$, {\it etc\/}. Notice the
similarity between the left-hand sides of \eqref{bo1}-\eqref{bo2} and
\eqref{oo1}-\eqref{oo2}. This arises as a consequence of the fact that
once the bosonic fluctuations
are gauge fixed, both they, and the Grassmann fluctuations, are
geometrically related to tangent vectors to the instanton moduli space
at the critical point.\footnote{More precisely, for each $A$ the Grassmann
collective coordinates are to {\it symplectic\/} tangent
vectors to the hyper-K\"ahler instanton moduli space: see Ref.~\cite{MY}.} 
We will use the ADHM constraints \eqref{bo1}-\eqref{bo2} and
\eqref{oo1}-\eqref{oo2} to eliminate the fluctuations $[a'_n]_{pq}$
and $[\CM^{\prime A}_\alpha]_{pq}$, $p,q=1,2$ and $2,1$. When this done a
non-trivial Jacobian factor of $Y^8$ results. Putting this together
with the Jacobian factor in \eqref{gfj} gives the function
\EQ{
Y^8\big(\zeta+Y^2\big)^2
\label{ntj}
}
that will be required later.

The next problem is to expand the 
the instanton effective action \eqref{jslmm} to Gaussian order in the
fluctuations. First the bosonic
pieces. To Gaussian order around the critical point we decompose
\EQ{
S_{\rm b}=S^{(1)}_{\rm b}+S^{(2)}_{\rm b}+S^{(12)}_{\rm b}\cdots\ ,
}
where $S^{(i)}_{\rm b}$ include all the terms that 
pertain separately to each of the topicons:
\EQ{
\frac1{4\pi^2}S^{(i)}_{\rm
b}=\zeta[\Bchi]_i^2+
\sum_{u=1\atop(\neq u_1,u_2)}^N
\Bphi_{uu_i}^2\big|w_{ui\aD}\big|^2\ ,
}
while the third term describes the interactions between the topicons:
\EQ{
\frac1{4\pi^2}S^{(12)}_{\rm
b}=\Bphi^2\big(1+\zeta/Y^2)
\big([\bar w^\aD]_{21}[w_{\aD}]_{12}+[\bar
w^\aD]_{12}[w_{\aD}]_{21}\big)
+2(\zeta+Y^2)[\Bchi]_{21}\cdot[\Bchi]_{12}\ ,
\label{act1}
}
where we have defined
\EQ{
\Bphi\equiv\Bphi_{u_1u_2}\ .
}

The fermionic part of the action has a similar decomposition at Gaussian order
\EQ{
S_{\rm f}=S_{\rm f}^{(1)}+S_{\rm f}^{(2)}+S^{(12)}_{\rm f}+\cdots\ ,
}
where
\EQ{
\frac1{2\pi^2}S_{\rm f}^{(i)}=-\tfrac18M_{AB}\xi_i^{\alpha
A}\xi^{B}_{i\alpha}+
\sum_{u=1\atop(\neq u_1,u_2)}^N\bar\mu^A_{iu}\big(
\bar\BSigma\cdot\Bphi_{uu_i}-M\big)_{AB}
\mu_{ui}^B
\label{looki}
}
and 
\SP{
\frac1{2\pi^2}S_{\rm
f}^{(12)}&=[\bar\mu^A]_{21}(\phi-M)_{AB}[\mu^B]_{12}
-[\mu^A]_{21}(\phi-M)_{AB}[\bar\mu^B]_{12}\\
&+[\CM^{\prime\alpha A}]_{21}
(\phi-M)_{AB}[\CM^{\prime B}_\alpha]_{12}
+[\CM^{\prime\alpha A}]_{21}[\chi_{AB}]_{12}\xi^B_\alpha
+\xi^{\alpha A}[\chi_{AB}]_{21}[\CM^{\prime B}_\alpha]_{12}\ .
\label{look}
}
In the above, and in much of the following, we use the notation
\EQ{
\phi_{AB}\equiv\bar\BSigma_{AB}\cdot\Bphi\
,\qquad\chi_{AB}\equiv\bar\BSigma_{AB}\cdot\Bchi\ .
}
The two sets of relative 
supersymmetric modes $\xi^A_\alpha$ with $A=\hat A=3,4$ and
$A=A'=1,2$, respectively, are treated differently. 
The integrals over 
$\xi_\alpha^{\hat A}$ are saturated by the mass terms in \eqref{looki},
whereas those over $\xi^{A'}_\alpha$ are saturated by interactions with
the bosonic fluctuations in \eqref{look}.
The integrals over the fluctuations $[\mu^A]_{pq}$ and
$[\bar\mu^A]_{pq}$, $p\neq q$, are simplified by shifting them by appropriate
amounts of $\xi^{A'}_\alpha$ in order to ``complete the square''. This
gives the equivalent form
\SP{
\frac1{2\pi^2}S_{\rm
f}^{(12)}&=\big(1+\zeta/Y^2\big)\Big\{[\bar\mu^A]_{21}(\phi-M)_{AB}[\mu^B]_{12}
-[\mu^A]_{21}(\phi-M)_{AB}[\bar\mu^B]_{12}\Big\}\\
&+(\zeta+Y^2)^{-1}\xi^{\alpha A'}
\Big\{\phi_{A'B'}[a'_{\alpha\aD}]_{21}
[\bar a^{\prime\aD\beta}]_{12}-[a'_{\alpha\aD}]_{21}\bar Y^{\aD\beta}
[\chi_{A'B'}]_{12}\\
&-[\chi_{A'B'}]_{21}Y_{\alpha\aD}[\bar 
a^{\prime\aD\beta}]_{12}-\zeta[\chi_{A'C}]_{21}(\phi-M)^{-1CD}[\chi_{DB'}]_{12}
\Big\}\xi^{B'}_\beta\ .
\label{look2}
}
In \eqref{look2} we should substitute for $[a'_{\alpha\aD}]_{pq}$ by
using the ADHM constraints \eqref{bo1}-\eqref{bo2}. 

Now we begin to integrate. First of all, the integrals over
$w_{u1\aD}$, $\mu^A_{u1}$, $\bar\mu^A_{1u}$, for $u\neq u_2$,
and $[\Bchi]_{1}$ 
completely decouple from the remaining integrals (and similarly for
$1\leftrightarrow2$). These integrals are
identical to the single instanton integrals done in Section 3. What
results is the non-trivial factor
\EQ{
\prod_{i=1}^2\prod_{{u=1\atop(\neq
u_1,u_2)}}^N\Big(1-\frac{m^2}{\phi_{uu_i}^2}\Big)\ .
\label{uuy}
}

Now we describe the remaining integrals over the fluctuations the
couple the two topicons. The integrals over the Grassmann
fluctuations $[\mu^A]_{pq}$ and $[\bar\mu^A]_{pq}$ produces the
factors\footnote{In the following we shall not indicate the
appropriate multiplicative 
numerical factors but simply collect them in the final 
expression.}
\EQ{
\big(1+\zeta/Y^2\big)^8\phi^{*4}(m^2-\phi^2)^2\ .
\label{phey}
}
Next we integrate over the relative supersymmetric modes
$\xi^A_\alpha$. The four $\xi^{\hat A}_\alpha$ are saturated by the
mass terms in \eqref{looki} leaving the four $\xi^{A'}_\alpha$ to be
saturated by the interactions in \eqref{look2}. Integrating out these
latter four variables produces a number of terms, however, many of
them involve odd functions of some of the components of the
six-vector $[\Bchi]_{pq}$
and will subsequently integrate to zero. The only terms which 
subsequently lead to non-zero contributions are
\SP{
&\zeta^2(\zeta+Y^2)^{-2}\bigg\{Y^{-4}\phi^{*2}\Big(\big([\bar w^\aD]_{21}[ 
w_{\aD}]_{12}+[\bar w^\aD]_{12}[ 
w_{\aD}]_{21}\big)^2-4[\bar w^\aD]_{12}[ 
w_{\aD}]_{12}[\bar w^\aD]_{21}[ 
w_{\aD}]_{21}\Big)\\
&-\frac{4m^2}{(m^2-\phi^2)^2}\Big(\sum_{a=3}^6[\chi_a]_{21}^2
[\chi_a]_{12}^2+2[\chi_3]_{21}[\chi_5]_{21}[\chi_3]_{12}[\chi_5]_{12}+
2[\chi_4]_{21}[\chi_6]_{21}[\chi_4]_{12}[\chi_6]_{12}\Big)\\
&\qquad\qquad\qquad-\frac{4\phi^2}{(m^2-\phi^2)^2}\sum_{a,b=3\atop(a\neq
b)}^6[\chi_a]_{21}[\chi_b]_{21}[\chi_a]_{12}[\chi_b]_{12}\bigg\}\ .
\label{hdd}
}
Now we turn to the bosonic fluctuations $[w_\aD]_{pq}$ and
$[\Bchi]_{pq}$, $p,q=1,2$ and $2,1$. The integrals involve \eqref{hdd} as an
insertion into the Gaussian integrals. The result, when amalgamated
with the non-trivial factors in Eqs.~\eqref{ntj} and \eqref{phey}, is 
\EQ{
m^2\Big(1-\frac{m^2}{\phi^2}\Big)^2
\Big[\frac1{\phi^2}-\frac1{2(\phi+m)^2}
-\frac1{2(\phi-m)^2}\Big]\frac{\zeta^2}{(\zeta+Y^2)^4}\ .
}
We can now
integrate over the relative position of the topicons:
\EQ{
\int d^4Y\,\frac{\zeta^2}{(\zeta+Y^2)^4}=\frac{\pi^2}{6}\ .
}

Finally putting all the non-trivial factors together
with the correct numerical factors
and restoring the notation $\phi\equiv\phi_{u_1u_2}$,
gives the final contribution of the critical-point set to the
centred instanton partition function
\EQ{
2m^2\Big(1-\frac{m^2}{\phi_{u_1u_2}^2}\Big)^2
\Big[\frac1{\phi_{u_1u_2}^2}-\frac1{2(\phi_{u_1u_2}+m)^2}
-\frac1{2(\phi_{u_1u_2}-m)^2}\Big]
\prod_{i=1}^2\prod_{{u=1\atop(\neq
u_1,u_2)}}^N\Big(1-\frac{m^2}{\phi_{uu_i}^2}\Big)\ .
}
Notice that the result is holomorphic in the VEVs and independent of
$\zeta^c$ as expected by our general cohomological argument.
Summing over the $\tfrac12N(N-1)$ critical-point sets of this type gives
the following contribution to the partition function:
\EQ{
m^2\sum_{u,v=1\atop
(u\neq
v)}^NT_u(\phi_u)T_v(\phi_v)
\Big[\frac1{\phi_{uv}^2}-\frac1{2(\phi_{uv}-m)^2}
-\frac1{2(\phi_{uv}+m)^2}\Big]\ ,
\label{res2}
}
where we have written the answer in terms of the functions $T_u(x)$
defined in \eqref{deft}.

\subsection{Topicons of the same flavour}

There are $N$ critical-points describing two topicons of the
same flavour $u_1=u_2\equiv
v\in\{1,\ldots,N\}$. On the critical submanifold  $\{w_{vi\aD},a'_n\}$
and $\{\mu^A_{vi},\bar\mu^A_{iv},\CM^{\prime
A}_\alpha\}$ satisfy the ADHM constraints, \eqref{adhm} and
\eqref{fadhm}, respectively, of two
instantons in a non-commutative $\U(1)$ theory. In other words the
critical submanifold is simply
\EQ{
\widehat\ms_{2,1}^{(\zeta)}\ .
}
The remaining
variables all vanish and are treated as fluctuations.

As previously, it is convenient to
shift the auxiliary variable $\Bchi$ by its critical-point value:
\EQ{
\Bchi\to\Bchi-\Bphi_v1_{\sst[2]\times[2]}\ .
\label{shift}
}
We now expand in the fluctuations $\{w_{ui\aD},\mu^A_{ui},
\bar\mu^A_{iu}\}$, for $u\neq v$. Since all the components of the
ADHM constraints are non-trivial at leading order the fluctuations
decouple from the $\delta$-functions which impose the constraints
\eqref{adhm} and \eqref{fadhm}.
The fluctuation integrals only involve
the integrand $\exp(-S-mS_{\text{mass}})$, 
where the action is expanded to Gaussian order around
the critical submanifold.
However, it is important, as we shall see below, to leave $\Bchi$
arbitrary rather than set it to its critical-point value; namely,
$\Bchi=0$ (after the shift \eqref{shift}). The fluctuation 
integrals produce the non-trivial factor
\EQ{
{\cal G}(\chi)\equiv\prod_{u=1\atop(\neq v)}^N
{\rm
det}_2\Big(1_{\sst[2]\times[2]}-m^2(\chi+\phi_{uv}
1_{\sst[2]\times[2]})^{-2}\Big) 
=T_v(\phi_v-\lambda_1)T_v(\phi_v-\lambda_2)\ .
\label{above}
}
Here, $\lambda_i$, $i=1,2$, are the eigenvalues of the $2\times2$
matrix $\chi$ and $T_u(x)$ was defined in \eqref{deft}.

The remaining integral is of the form
\EQ{
\int_{\widehat\ms_{2,1}}
\Bomega^{\sst(\N=4)}\,e^{-S-mS_{\text{mass}}}{\cal G}(\chi)\ .
\label{uup}
}
where ${\cal G}(\chi)$ is the non-trivial function \eqref{above}.
There are two
types of contribution depending on whether the mass term
$S_{\text{mass}}$ is employed to saturate any Grassmann integrals. 
For the first $S_{\text{mass}}$ is not used and then it is easy to see
that the $\SU(4)$ symmetry of the resulting integral means that any
insertion of powers of $\lambda_i$ integrate to zero. Hence, only the
value of ${\cal
G}(\chi)$ at $\chi=0$ contributes.
The second occurs when the mass terms are used
to saturate the integrals over the four Grassmann collective coordinates
left over from the 
set of eight $\{\mu^{\hat A},\bar\mu^{\hat A},\CM^{\prime \hat
A}_\alpha\}$ once 
the ADHM constraints \eqref{fadhm} have been imposed. What is left is the
volume form of the $\N=2$ theory since, when the coordinates
$\{\mu^{\hat A},\bar\mu^{\hat A},\CM^{\prime \hat A}_\alpha\}$ 
are set to zero, 
the instanton effective action 
reduces to that of the $\N=2$ theory denoted $S^{\sst(\N=2)}$.
This integral is precisely the same
as the one that appeared in Ref.~\cite{local} where we argued that 
only terms quadratic
in the expansion of ${\cal G}(\chi)$ contribute. Hence, 
\eqref{uup} is
\SP{
&\int_{\widehat\ms_{2,1}}
\Bomega^{\sst(\N=4)}\,e^{-S-mS_{\text{mass}}}{\cal G}(\chi)\\
&=T_v(\phi_v)^2\int_{\widehat\ms_{2,1}}
\Bomega^{\sst(\N=4)}\,e^{-S}
+m^2\left(\frac{\partial
T_v(\phi_v)}{\partial\phi_v}\right)^2\int_{\widehat\ms_{2,1}}
\Bomega^{\sst(\N=2)}\,e^{-S^{\sst(\N=2)}}
\lambda_1\lambda_2\\
&\qquad\qquad\qquad+\tfrac12m^2T_v(\phi_v)\frac{\partial^2T_v(\phi_v)}
{\partial\phi_v^2}\int_{\widehat\ms_{2,1}}
\Bomega^{\sst(\N=2)}\,e^{-S^{\sst(\N=2)}}
(\lambda_1^2+\lambda_2^2)\ .
}
In the above, the first term here involves the centred instanton
partition on the non-commutative $\U(1)$ 
two-instanton moduli space. This integral can be
evaluated explicitly using the formulae of the Appendix in
Ref.~\cite{local}. However, from our earlier discussion 
we also know that this partition function is precisely the 
Gauss-Bonnet-Chern integral on $\widehat\ms_{2,1}^{(\zeta)}$
\cite{MY} and this four-dimensional space is the Eguchi-Hanson
manifold \cite{Lee:2001hp} whose Gauss-Bonnet-Chern integral is well
known to be $\tfrac32$ \cite{EGH}. 
The other terms were evaluated in \cite{local}:
\EQ{
\int_{\widehat\ms_{2,1}}
\Bomega^{\sst(\N=2)}\,e^{-S^{\sst(\N=2)}}\lambda_1\lambda_2=0\ ,\qquad
\int_{\widehat\ms_{2,1}}
\Bomega^{\sst(\N=2)}\,e^{-S^{\sst(\N=2)}}
(\lambda_1^2+\lambda_2^2)=\frac12\ . 
}
Hence, the final result for the contributions from two
topicons of the same flavour to 
the centred instanton partition function is
\EQ{
\sum_{u=1}^N\Big(\tfrac32T_u(\phi_u)^2+\tfrac14m^2T_u(\phi_u)
\frac{\partial^2T_u(\phi_u)}{\partial\phi_u^2}\Big)\ .
\label{res1}
}

Finally, summing \eqref{res1} and \eqref{res2} we have the
centred two-instanton partition function
\SP{
\widehat{\EuScript Z}_{2,N}^{\sst(\N=2^*)}&=\sum_{u=1}^N
\Big(\tfrac32T_u(\phi_u)^2+\tfrac14m^2T_u(\phi_u)
\frac{\partial^2T_u(\phi_u)}{\partial\phi_u^2}\Big)\\
&+m^2\sum_{u,v=1\atop(u\neq
v)}^NT_u(\phi_u)T_v(\phi_v)\Big[\frac1{\phi_{uv}^2}-\frac1{2(\phi_{uv}-m)^2}
-\frac1{2(\phi_{uv}+m)^2}\Big]\ .
\label{scon}
}
Using the relation \eqref{ndeq}, we find precisely the Seiberg-Witten
prediction for the $k=2$ instanton coefficient of the prepotential
\eqref{ktwo}. 

\section{Arbitrary Instanton Number}

Now we proceed to the wholly more ambitious proposition of calculating
the prepotential for instanton charge $k>2$. In fact we shall 
show that it is possible to calculate the terms of ${\cal O}(m^4)$ in
the prepotential for all
instanton charge. These are the first non-trivial, {\it
i.e.\/}~VEV-dependent, terms in the expansion of the instanton portion
of the prepotential in $m^2$. The prediction from Seiberg-Witten
theory follows from Ref.~\cite{MNW} and is quoted in \eqref{impp}. The
verification of this prediction will provide by far the most stringent test
of Seiberg-Witten theory to date.

The key step in evaluating the ${\cal O}(m^4)$ contribution to ${\cal
F}_k$ is the following:

{\bf Proposition:}
{\it The ${\cal O}(m^2)$ contribution to
$\widehat\Z_{k,N}^{\sst(\N=2^*)}$ comes
exclusively from $k$ topicons of the same flavour.}

{\bf Proof:}
First of all, it is easy to see that if the component of the
critical-point set corresponds to a partition with at least 3 non-trivial
blocks, {\it i.e.\/}~some $k_{u_1},k_{u_2},k_{u_3},\ldots, k_{u_p}>0$,
with $p>2$, then
the contribution must be at least ${\cal O}(m^6)$. The reason is that
the supersymmetric modes associated to each flavour of topicon,
${\rm tr}_{k_{u_\ell}}\CM^{\prime \hat A}_\alpha$,
$\ell=1,\ldots,p$, number in total $4(p-1)$.\footnote{The traces
here are taken in each block and the $-4$
arises because of the overall traceless condition which implies
$\sum_{\ell}{\rm tr}_{k_{u_\ell}}\CM^{\prime \hat A}_\alpha=0$.}
The integrals over these variables must be saturated by mass terms,
giving a factor of $m^{2(p-1)}$ in the contribution to
$\widehat\Z_{k,N}^{\sst(\N=2^*)}$. So for $p>2$ these give at least
${\cal O}(m^6)$ contributions to the prepotential.

The argument above fails when there are precisely two blocks, {\it
i.e.\/}~just two flavours of topicons, and this
case must be considered more explicitly.
We have already seen by explicit computation in the
case $k=2$ in Section 4.1, and in particular
\eqref{res2}, that this contribution is actually ${\cal O}(m^4)$ in
partition function and so ${\cal O}(m^6)$ in the
prepotential. However, when one examines the reason for this it is
because the term in square brackets in \eqref{res2}
vanishes like ${\cal O}(m^2)$. In other words, in order to see the
result relies on a delicate
cancellation between terms. So we
must prove something similar occurs in a partition $k\to k_1+k_2$ with
$k>2$ and this, unfortunately, requires us to 
perform the integrals over the fluctuations explicitly.

First of all, let us separate out the variables
$\{w_{ui\aD},\mu^A_{ui},\bar\mu^A_{iu}\}$, $u\neq u_1,u_2$. As we have
seen in Section 4.1, the integrals over these fluctuations 
decouples from the the rest and they will play no r\^ole in the
following argument. The remaining variables then have an obvious 
$2\times2$ block form, where the blocks are of size $k_1$ and
$k_2$. We will indicate the diagonal block by $[a'_n]_p$, $p=1,2$,
and the off-diagonal blocks by 
$[a'_n]_{pq}$, $p,q=1,2$ and $2,1$, 
{\it etc\/}.\footnote{Note that $[a'_n]_{pq}$ is a
$k_p\times k_q$ matrix, while $[w_\aD]_{pq}$ is a $1\times k_q$
matrix, {\it etc\/}.}
The two block-diagonal elements satisfy the ADHM constraints of a
charge $k_1$, respectively $k_2$, $\U(1)$ instanton which parameterize
the critical submanifold. The off-diagonal
variables are the treated as fluctuations. 

Each of the diagonal blocks
has 8 supersymmetric modes. However, since we are considering the centred
instanton moduli space only the 8 relative supersymmetric modes
defined by
\EQ{
\CM^{\prime A}_\alpha=\frac1k\MAT{k_21_{\sst[k_1]\times[k_1]}&0\\
0&-k_11_{\sst[k_2]\times[k_2]}}\xi^A_\alpha
\label{relsm}
}
are relevant.
At leading order around the critical submanifold, 
the integrals over the
four modes $\xi^{\hat A}_\alpha$ must be lifted by the mass terms
$S_{\text{mass}}$ yielding a factor of $m^2$. Hence we must show that
the remaining integral vanishes at ${\cal O}(m^0)$ in order to
complete the proof. We now focus on the other four relative supersymmetric
modes $\xi^{A'}_\alpha$. The integrals over these variables must be
lifted via interactions with the fluctuations as we found in the case
$k_1=k_2=1$ in Section 4.1. In fact the following analysis is very
similar to that in Section 4.1 but with the added complication that we
have to keep track of the additional non-trivial matrix structure for
$k_1,k_2>1$. 

First of all, consider the issue
of gauge fixing. As in 
Section 4.1 we will fix the gauge by demanding that the
fluctuations are orthogonal to $\U(k)$ transformations acting on the
critical-point 
solution. This is convenient because then, as we saw in Section 4.1, 
the ADHM constraints along
with the extra gauge-fixing constraint imply that the bosonic
fluctuations satisfy the fermionic ADHM constraints at the critical
point. The gauge-fixing conditions are
\EQ{
[\bar w^\aD]_{pq}[w_\aD]_{q}+[\bar w_\aD]_{p}[w^\aD]_{pq}
+[\bar
a^{\prime\aD\alpha}]_{pq}[a'_{\alpha\aD}]_{q}
-[a'_{\alpha\aD}]_{p}[\bar
a^{\prime\aD\alpha}]_{pq}=0\ .
}
Hence, the off-diagonal ADHM constraints plus
gauge-fixed conditions are
\EQ{
[\bar w^\aD]_{pq}[w_\bD]_{q}+[\bar w_\bD]_{p}
[w^\aD]_{pq}+[\bar
a^{\prime\aD\alpha}]_{pq}[a'_{\alpha\bD}]_{q}
-[a'_{\alpha\bD}]_{p}[\bar
a^{\prime\aD\alpha}]_{pq}=0\ ,
\label{iut}
}
with no sum on $p,q$. 
Similarly the off-diagonal 
fermionic ADHM constraints are
\SP{
&[\bar\mu^A]_{pq}[w_\bD]_{q}+[\bar w_\bD]_{p}
[\mu^A]_{pq}+[\CM^{\prime \alpha A}]_{pq}[a'_{\alpha\bD}]_{q}
-[a^{\prime}_{\alpha\bD}]_{p}[\CM^{\prime\alpha A}]_{pq}\\
&=
-[\bar\mu^A]_{p}[w_\bD]_{pq}-[\bar w_\bD]_{pq}
[\mu^A]_{q}-[\CM^{\prime \alpha A}]_{p}[a'_{\alpha\bD}]_{pq}
+[a^{\prime}_{\alpha\bD}]_{pq}[\CM^{\prime\alpha A}]_{q}\ .
\label{odfadhm}
}
Note the similarity between the left-hand sides of \eqref{iut} and
\eqref{odfadhm}. 

The total number of constraints \eqref{iut} and \eqref{odfadhm} is
$8k_1k_2$ and $16k_1k_2$ respectively, matching precisely the number for the 
fluctuations $[a'_n]_{pq}$ or
$[\CM^{\prime A}]_{pq}$, $p,q=1,2$ and $2,1$. Hence, we can use the
constraints to eliminate these variables.
The action for the fluctuations follows by
expanding \eqref{jslmm} to Gaussian order. In the bosonic sector, the
relevant terms are
\EQ{
\frac1{4\pi^2}S^{(12)}_{\text{b}}=
{\rm tr}\bigg\{\Bphi^2\MAT{\bar w^\aD&, & -w^\aD}_{21}\BDelta
\MAT{w_\aD\\
\bar w_\aD}_{12}
+2[\Bchi]_{21}\cdot\BOmega[\Bchi]_{12}\bigg\}\ .
\label{bosa}
}
where we have defined the following linear operators which depend on
the critical-point topicon collective coordinates
\SP{
\AA^\bD\MAT{w_\aD\\
\bar w_\aD}_{12}&=[\bar w^\bD]_{1}[w_\aD]_{12}+
[\bar w_\aD]_{12}[w^\bD]_{2}\ ,\\
\MAT{\bar w^\aD & , &
-w^\aD}_{21}\BB_\bD&=[\bar w_\bD]_{2}[w^\aD]_{21}+
[\bar w^\aD]_{21}[w_\bD]_{1}\ ,\\
\UU_{\alpha\bD}\Theta&=\Theta[a'_{\alpha\bD}]_{2}-
[a'_{\alpha\bD}]_{1}\Theta\ ,
}
in terms of which\footnote{To connect with Section 4.1, in the case
$k_1=k_2=1$ we have $\BOmega=\zeta+Y^2$ and $\BDelta=1+\zeta/Y^2$.}
\EQ{
\BOmega=\tfrac12\big(\AA^\aD\BB_\aD+\bar
\UU^{\aD\alpha}\UU_{\alpha\aD}\big)\ ,\qquad
\BDelta=1+\BB_\aD\bar\UU^{-1\aD\alpha}\UU^{-1}_{\alpha\bD} \AA^\bD\ .
}
In the above, $\Theta$ is an arbitrary 
$k_1\times k_2$ matrix. The resulting machinations are 
greatly simplified by noting that for an arbitrary $k_1\times k_2$
matrix $\Theta$ 
\SP{
\big(\AA^\aD\BB_\bD+\bar\UU^{\aD\alpha}\UU_{\alpha\bD}\big)\Theta
&=[\bar w^\aD w_\bD+\bar
a^{\prime\aD\alpha}a'_{\alpha\bD}]_{1}
\Theta-\Theta[\bar w_\bD w^\aD-
a'_{\alpha\bD}\bar a^{\prime\aD\alpha}]_{2}\\
&\qquad\qquad\qquad-[\bar
a^{\prime\aD\alpha}]_{1}\Theta[a'_{\alpha\bD}]_{2}-
[
a'_{\alpha\bD}]_{1}\Theta[\bar a^{\prime\aD\alpha}]_{2}\\
&
=\delta^\aD{}_\bD\Big([\sigma]_1\Theta+\Theta[\sigma]_2-2
[a'_n]_{1}\Theta[a'_n]_{2}\Big)\ ,
\label{bbc}
}
where we employed the ADHM constraints of the critical-point solutions:
\EQ{
[\bar w^\aD w_\bD+\bar
a^{\prime\aD\alpha}a'_{\alpha\bD}]_{p}=\tfrac12\zeta^c\tau^{c\aD}{}_\bD
+[\sigma]_{p}\delta^\aD{}_\bD\ .
}
which defines the $k_p\times k_p$ matrices $[\sigma]_{p}$. 
In addition, \eqref{bbc} implies
\EQ{
\BOmega\Theta=[\sigma]_{1}\Theta+\Theta[\sigma]_{2}-2
[a'_n]_{1}\Theta[a'_n]_{2}\ .
}

Now we turn to the Grassmann fluctuations. The fermionic ADHM
constraints \eqref{odfadhm} 
are used to eliminate the fluctuations $[\CM^{\prime
A}_\alpha]_{pq}$. Notice, the right-hand sides of these constraints
involves the Grassmann collective coordinates of the critical-point
solution. To leading order in the mass we can ignore the
coupling of the fluctuations to all but the relative supersymmetric
modes $\xi^{A'}_\alpha$ in \eqref{relsm}. The reason is that these
modes are not lifted by any other effects, while using the constraints
to lift other modes of the critical-point solution inevitably ends up costing
powers of $m$. At leading order in our mass expansion we can also set
$m=0$ in the action of the fluctuations:
\SP{
\frac1{2\pi^2}S_{\text{f}}^{(12)}
&={\rm tr}\bigg\{\MAT{
\bar\mu^A& , & -\mu^A}_{21}\phi_{AB}\MAT{\mu^B\\
\bar\mu^B}_{12}\\ &+ \Big[\MAT{
\bar\mu^A & , & -\mu^A}_{21}\BB_\aD\bar\UU^{-1\aD\alpha}-\xi^{\gamma
A'}[a'_{\gamma\aD}]_{21}\bar\UU^{-1\aD\alpha}
+\xi^{\alpha C'}[\chi_{C'D}]_{21}\phi^{-1DA}\Big]\phi_{AB}\\
&\qquad\qquad\times
\Big[\UU^{-1}_{\alpha\bD}\AA^\bD\MAT{\mu^B
\\ \bar\mu^B}_{12}-\UU^{-1}_{\alpha\bD}
[\bar
a^{\prime\bD\delta}]_{12}
\xi^{B'}_\delta+\phi^{-1BE}[\chi_{EF'}]_{12}
\xi_{\alpha}^{F'}\Big]\\
&\qquad\qquad\qquad-\xi^{\alpha A'}[\chi_{A'C}]_{21}\phi^{-1CD}[\chi_{DB'}]_{12}
\xi_{\alpha}^{B'}\bigg\}\ .
\label{poo}
}
Now we shift the fluctuations $[\mu^A]_{pq}$ and
$[\bar\mu^A]_{pq}$ by 
appropriate amounts of $\xi^{A'}_\alpha$ in order to ``complete the
square''. This leaves the terms which are responsible for the lifting
the relative supersymmetric modes $\xi^{A'}_\alpha$. The expression involves a
quadratic interaction of $\xi^{A'}_\alpha$:
\SP{
&\frac1{4\pi^2}S_{\text{int}}=\xi^{\alpha A'}{\rm tr}\bigg\{\phi_{A'B'}
[a'_{\alpha\aD}]_{21}\BOmega^{-1}[\bar a^{\prime\aD\beta}]_{12}-
[a'_{\alpha\aD}]_{21}\BOmega^{-1}\bar\UU^{\aD\beta}[\chi_{A'B'}]_{12}\\
&-
[\chi_{A'B'}]_{21}\UU_{\alpha\aD}\BOmega^{-1}
[\bar a^{\prime\aD\beta}]_{12}
-\delta_\alpha{}^\beta[\chi_{A'C}]_{21}\phi^{-1CD}\Big(
1-\UU_n\BOmega^{-1}\UU_n\Big) 
[\chi_{DB'}]_{12}\bigg\}\xi^{B'}_\beta\ .
\label{qer}
}
In our conventions
\EQ{
\UU_{\alpha\aD}=\UU_n\sigma_{n\alpha\aD}\ ,\qquad
\bar\UU^{\aD\alpha}=\UU_n\bar\sigma_n^{\aD\alpha}\ ,\qquad
\UU^{-1}_{\alpha\aD}\bar\UU^{\aD\beta}=\delta_\alpha{}^\beta\ ,\qquad
\bar\UU^{-1\aD\alpha}\UU_{\alpha\bD}=\delta^\aD{}_\bD\ .
}

The integrals over the four relative supersymmetric modes are saturated
by pulling down two powers of the action \eqref{qer}. This yields an
expression which is 
quartic in the bosonic fluctuations. The constraints 
\eqref{iut} are used to eliminate $[a'_{\alpha\aD}]_{pq}$ in
favour of $[w_\aD]_{pq}$. The remaining expression is rather
complicated, however, many of the terms will subsequently integrate to
zero since they are odd functions in some of the components of the 
bosonic fluctuations.
Ignoring these terms, the relevant part of the integrand is
\SP{
&\phi^{*2}\big({\cal F}_\beta{}^\alpha{\cal F}_\alpha{}^\beta+
{\cal F}_\beta{}^\alpha{\cal F}^\beta{}_\alpha\big)\\
&-\frac2{\phi^2}
\sum_{a,b=3\atop(a\neq b)}^6{\rm
tr}\Big\{[\chi_a]_{21}\big(1
-\UU_n\BOmega^{-1}\UU_n\big)[\chi_b]_{12}\Big\}{\rm tr}
\Big\{[\chi_b]_{21}\big(1
-\UU_n\BOmega^{-1}\UU_n\big)[\chi_a]_{12}\Big\}+\cdots\ , 
\label{xtr}
}
where
\EQ{
{\cal F}_\beta{}^\alpha={\rm tr}\bigg\{\MAT{\bar w^\aD & , &
-w^\aD}_{21}
\BB_\bD\bar\UU^{-1\bD\alpha}\BOmega^{-1}\UU^{-1}_{\beta\gD}\AA^\gD
\MAT{w_\aD\\ \bar w_\aD}_{12}\bigg\}\ .
}

What remains is to integrate over the 
Gaussian fluctuations $[w]_{pq}$ and
$[\chi]_{pq}$, which are governed by the action \eqref{bosa}, with
with \eqref{xtr}, a quartic function of the fluctuations, as 
an integrand. Let us concentrate on the first term in \eqref{xtr},
quartic in the fluctuations $[w]_{pq}$. These variables have a
``propagator'' 
\EQ{
\frac1{\Bphi^2}\BDelta^{-1} \delta_\aD{}^\bD\ .
}
Performing the Gaussian
integrals yields two types of term: the first consisting of a 
product of two ``trace'' factors and a second with a single
trace. Let us dispense with the first. The contributions are of the
form of the first set of terms in \eqref{xtr} with
\EQ{
{\cal F}_\beta{}^\alpha\longrightarrow\frac1{\Bphi^2}{\rm tr}\Big(
\BB_\aD\UU^{-1\aD\alpha}\BOmega^{-1}\bar\UU^{-1}_{\beta\bD}\AA^\bD\BDelta^{-1}
\Big)\ .
\label{guu}
}
Now we use the identity
\EQ{
\UU^{-1}_{\beta\gD}\AA^\gD\BDelta^{-1}
\BB_\bD\bar\UU^{-1\bD\alpha}=\delta_\beta{}^\alpha-\tfrac12\UU_{\beta\aD}
\BOmega^{-1}\bar\UU^{\aD\alpha}=
\delta_\beta{}^\alpha\big(1
-\UU_n\BOmega^{-1}\UU_n\big)
\label{did}
}
to prove that the right-hand of \eqref{guu} is proportional to
$\delta_\beta{}^\alpha$. 
Given this and the combination of ${\cal F}_\beta{}^\alpha$ in \eqref{xtr},
we see that the double trace terms do not
contribute. The single trace contribution is
\SP{
&\frac1{\phi^2}
{\rm tr}\Big(
\BB_\aD
\bar\UU^{-1\aD\alpha}\BOmega^{-1}\UU^{-1}_{\beta\bD}\AA^\bD\BDelta^{-1}
\BB_\gD\bar\UU^{-1\gD\beta}\BOmega^{-1}\UU^{-1}_{\alpha\dD}\AA^\gD\BDelta^{-1}
\\
&\qquad\qquad+
\BB_\aD\bar\UU^{-1\aD\alpha}\BOmega^{-1}\UU^{-1}_{\beta\bD}\AA^\bD\BDelta^{-1}
\BB_\gD\bar\UU^{-1\gD}{}_{\alpha}
\BOmega^{-1}\UU^{-1}{}^\beta{}_{\dD}\AA^\gD\BDelta^{-1}
\Big)\\
&\qquad\qquad=\frac6{\phi^2}{\rm tr}\Big\{\big(1
-\UU_n\BOmega^{-1}\UU_n\big)\BOmega^{-1}\big(1
-\UU_n\BOmega^{-1}\UU_n\big)
\BOmega^{-1}\Big\}
\ .
\label{grb}
}
where in the second line we used the identity \eqref{did}.

Now we turn to the contribution coming from the term quartic in
$[\chi_a]_{pq}$ in \eqref{xtr}. From \eqref{bosa}, the ``propagator''
for the $[\chi_a]_{pq}$ fluctuations is
\EQ{
\tfrac12\delta_{ab}\BOmega^{-1}\ .
}
Hence the contribution from these terms has the form
\EQ{
-\frac6{\phi^2}{\rm tr}\Big\{\big(1
-\UU_n\BOmega^{-1}\UU_n\big)\BOmega^{-1}\big(1
-\UU_n\BOmega^{-1}\UU_n\big)\BOmega^{-1}\Big\}\ .
\label{gra}
}
Remarkably \eqref{grb} precisely cancels the contribution from
\eqref{gra}, generalizing the vanishing of the square bracket in
\eqref{res2} for $m=0$. This completes the proof that the contribution
from the critical submanifolds
$\ms_{k_1,1}^{(\zeta)}\times\ms_{k_2,1}^{(\zeta)}/{\mathbb R}^4$
to the partition function vanishes to ${\cal
O}(m^2)$. \hspace*{\fill}{\it QED\/} 

Finally we consider the $N$ cases when the critical submanifold is
$\widehat\ms_{k,1}^{(\zeta)}$ corresponding to $k$ topicons all
of the same flavour. This is the situation considered in
Section 4.2 for $k=2$ and the generalization to $k>2$ is
reasonably straightforward. We focus on topicons of flavour $v$. 
Integrating over the
fluctuations $\{w_{ui\aD},\mu^A_{ui},\bar\mu^A_{iu}\}$, $u\neq v$
produces the factor 
\EQ{
{\cal G}(\chi)=\prod_{i=1}^kT_v(\phi_v-\lambda_i)\ ,
\label{defgg}
}
where $\lambda_i$ are the eigenvalues of the $k\times k$ matrix
$\chi$. The remaining integral is then 
\EQ{
\int_{\widehat\ms_{k,1}}
\Bomega^{\sst(\N=4)}\,e^{-S-mS_{\text{mass}}}{\cal G}(\chi)\
. 
\label{uupp}
}
We can now expand the factor $\exp-mS_{\text{mass}}$ in powers of the
mass. Due to the counting of fermion zero modes, 
non-trivial terms arise when only even powers of this term are
pulled down. A term of order $m^{2p}$ carries with it $4p$ factors of the
Grassmann collective coordinates indexed by $A=3,4$,
{\it i.e.\/}~collective coordinates in the set
$\{\mu^{\hat A},\bar\mu^{\hat A},\CM^{\prime \hat A}\}$. Consequently there is a mismatch
between the remaining Grassmann collective coordinates 
$\{\mu^{\hat A},\bar\mu^{\hat A},\CM^{\prime \hat A}\}$ and $\{\mu^{A'},\bar\mu^{A'},
\CM^{\prime A'}\}$. The excess of the latter set by $4p$ means that
after integrating them out the integrand will contain a factor of 
$(\chi^\dagger)^{2p}$ times an $\SU(4)$-invariant function of
$\Bchi$. Consequently in order to have a non-trivial integral we must
expand ${\cal G}(\chi)$ to $(2p)^{\rm th}$ order in $\chi$, or,
equivalently, its eigenvalues $\{\lambda_i\}$. 

To order $m^2$, 
we are interested in the first two terms in the expansion. The first is
simply
\EQ{
T_v(\phi_v)^k\int_{\widehat\ms_{k,1}}
\Bomega^{\sst(\N=4)}\,e^{-S}\ ,
\label{meo}
}
while the second term is of the form
\EQ{
\tfrac12m^2\sum_{i,j=1}^k
\frac{\partial^2{\cal
G}(\chi)}{\partial\lambda_i\partial\lambda_j}
\Big|_{\chi=0}
\int_{\widehat\ms_{k,1}}
\Bomega^{\sst(\N=4)}\,e^{-S}\big(S_{\text{mass}}\big)^2
\lambda_i\lambda_j\
.
}
However, from the form of ${\cal G}(\chi)$ in \eqref{defgg} one
readily shows that
\EQ{
\frac{\partial^2{\cal
G}(\chi)}{\partial\lambda_i\partial\lambda_j}
\Big|_{\chi=0}\ ,
}
is actually ${\cal O}(m^2)$ so that the only contribution at ${\cal
O}(m^2)$ comes from \eqref{meo} alone. The only mass dependence is
in the pre-factor and expanding to ${\cal O}(m^2)$ one finds
\EQ{
-km^2\sum_{v=1\atop(\neq u)}^N\frac1{(\phi_u-\phi_v)^2}
\int_{\widehat\ms_{k,1}}
\Bomega^{\sst(\N=4)}\,e^{-S}\ .
}
We recognize the integral over $\widehat\ms_{k,1}$ as the
Gauss-Bonnet-Chern integral of the resolved instanton moduli space. 
While this integral has not been
explicitly evaluated for all $k$, there are strong 
indirect arguments which give the general formula for
$k$-instantons \cite{Dorey:2001zq,Dorey:2001ym}:
\EQ{
\widehat\Z_{k,1}^{\sst(\N=4)}=\int_{\widehat\ms_{k,1}}
\Bomega^{\sst(\N=4)}\,e^{-S}=\sum_{d|k}\frac1d\ ,
\label{gbci}
}
where the sum is over the integer divisors of $k$. 

Summing over the $N$ types of topicon gives our final 
result for the partition function to ${\cal O}(m^2)$:
\EQ{
\widehat\Z_{k,N}^{\sst(\N=2^*)}=-m^2k\Big(\sum_{d|k}\frac1d\Big)
\sum_{u,v=1\atop(u\neq v)}^N\frac1{(\phi_u-\phi_v)^2}+{\cal O}(m^4)\ ,
}
modulo an irrelevant constant. Using
the fact that $k\sum_{d|k}d^{-1}=\sum_{d|k}d$, and employing
\eqref{ndeq}, we have for the ${\cal O}(m^4)$ contribution to the prepotential
\EQ{
{\cal F}_k=m^4\Big(\sum_{d|k}d\Big)
\sum_{u,v=1\atop(u\neq v)}^N\frac1{(\phi_u-\phi_v)^2}+{\cal O}(m^6)\ .
}
This matches the prediction from Seiberg-Witten theory \eqref{impp} exactly.

\section{Discussion}

We have already noted that our results 
are in precise agreement with the predictions coming from 
Seiberg-Witten Theory.
We can also compare the results of our calculation at the one-instanton
level with gauge group $\SU(2)$ with the brute-force integral
over the instanton moduli space in the commutative theory performed in
Ref.~\cite{DKMn4}. Written in our notation the result in that
reference is\footnote{In order to compare our formulae to those in
Ref.~\cite{DKMn4} replace $\phi\to\sqrt 2 v$
where $\phi=\phi_1-\phi_2$.} 
\EQ{
{\cal F}_1=-\frac{m^2}2+\frac{2m^4}{\phi^2}
\label{up1}
}
to compare with our result in the non-commutative theory (from
\eqref{ndeq} and \eqref{oneres})
\EQ{
{\cal F}_1^{\text{nc}}=-2m^2\Big(1-\frac{m^2}{\phi^2}\Big)\ .
\label{up2}
}
The mismatch between \eqref{up1} and \eqref{up2} can be understood in
precisely the way similar mismatches between the commutative and
non-commutative expressions for the prepotential were explained 
in \cite{local}. As
described in \cite{local} the integral over the resolved instanton
moduli space of the non-commutative theory misses contributions from
the singularities of the instanton moduli space in the commutative
theory. At the one-instanton level the contribution to the centred
instanton partition function from the
singularity is independent of the VEV and was calculated in 
Ref.~\cite{Dorey:2001zq} for arbitrary $N$:
\EQ{
{\cal S}_{1,N}=-\frac{2\Gamma(N+\tfrac12)}{\Gamma(N)\Gamma(\tfrac12)}\
,
}
so equal to $-\tfrac32$ for $N=2$: precisely accounting for the
mismatch between \eqref{up1} and \eqref{up2}. However, this mismatch
does not lead to any physical difference between the commutative and
non-commutative theories since it is independent of the VEV. 

The localization that we have described can be given a nice visual
interpretation involving D-branes in Type II string theory. It is
now well established that instantons in $\U(N)$ gauge theory
correspond to a certain decoupling limit of a configuration of
D-instantons in the vicinity of $N$ D3-branes. More generally it is
useful to
consider the D$p$-D$(p+4)$-brane system. The instanton calculus is
then obtained as the dimensional reduction of the theory on the
$k$ D$p$-branes which is some $\U(k)$ gauge theory that can be
formulated as a theory with eight supercharges
in a maximum dimension of six. This explains why the
$\N=4$ instanton partition function is the dimensional reduction of a gauged
linear $\sigma$-model in six dimensions with $\N=(1,0)$ supersymmetry
\cite{MY,MO3}. In general this
theory has a Higgs branch which corresponds to a situation where all the
D$p$-branes have been absorbed onto the D$(p+4)$-branes, 
a Coulomb branch where all the D$p$-branes move off into the bulk, as
well as various mixed branches describing intermediate
situations. To describe the Coulomb branch of the D$(p+4)$-brane
theory one separates these branes in their transverse space. On top of
this, the gauge theory on the D$(p+4)$-branes can be made
non-commutative by turning on certain background fields. The
background field act as Fayet-Illiopolos parameters in the D$p$-brane
theory which lifts the Coulomb and mixed branches leaving only the Higgs
branch: the D$p$-brane are forced onto the D$(p+4)$-branes. But
since the D$(p+4)$-branes are separated the D$p$-branes must choose
which out of the $N$ to be absorbed on. This is precisely the
picture that lies behind the combinatorics of the partitions in
\eqref{partitions}. The effective theory on each D$(p+4)$-brane is
then a non-commutative $\U(1)$ theory as suggested by the exact
component of the instanton moduli space in \eqref{partitions}. The
resulting $\U(1)$ integrals are still non-trivial because we have to
take account of interactions between D$p$-branes living on different
D$(p+4)$-branes as we have seen in our calculations.

\vspace{1cm}

I would like to thank Nick Dorey and Valya Khoze for many valuable
conversations that have contributed to this work and forced me to
sharpen up the concept of the topicon.

\startappendix

\Appendix{Conventions}

We frequently write a 4-vector $x_n$ as the $2\times2$ matrices 
\EQ{
x_{\alpha\aD}=x_n\sigma_{n\alpha\aD}\ ,\qquad
\bar x^{\aD\alpha}=x_n\bar\sigma_n^{\aD\alpha}\ ,
\label{rae}
}
where $\sigma_{n\alpha\aD}$ are the components of four $2\times2$ matrices
$\sigma_n=(i\vec\tau,1_{\sst[2]\times[2]})$, with
$\tau^c$, $c=1,2,3$ being the three Pauli
matrices, and  $\bar\sigma_n\equiv
\sigma_n^\dagger=(-i\vec\tau,1_{\sst[2]\times[2]})$
with components $\bar\sigma_n^{\aD\alpha}$. Spinor indices are raised
and lowered using the $\epsilon$-tensor as in \cite{WB}.

The $\N=4$ instanton calculus makes frequent use of the
$\Sigma$-matrices of $\SO(6)$ ($\simeq\SU(4)$). These can be thought
as Clebsch-Gordon coefficients which relate the spinor, anti-spinor
and vector representations. We think of $\BSigma^{AB}$ and
$\bar\BSigma_{AB}$ as six-vectors of $4\times4$ matrices. 
In our conventions we have
\SP{
\BSigma^{AB}&=\MAT{\a0& \phantom{-}1& \a0& \a0\\ -1& \a0& \a0& \a0\\ \a0& \a0& \a0& \phantom{-}1\\ 
\a0& \a0& -1& \a0}\ ,\quad  
  i\MAT{\a0& \phantom{-}1& \a0& \a0\\ -1& \a0& \a0& \a0\\ \a0& \a0& \a0& -1\\ \a0& \a0& \phantom{-}1& \a0}\ 
,\quad  \MAT{0& \a0& -1& 
      \a0\\ 0& \a0& \a0& \phantom{-}1\\ 1& \a0& \a0& \a0\\ 0& -1& \a0& \a0}\ ,\\
&\qquad\qquad  i\MAT{0& \a0& -1& \a0\\ 0& \a0& \a0& -1\\ 1& \a0& \a0& \a0\\ 0& \phantom{-}1& \a0& \a0}\ ,\quad 
  \MAT{\a0& \a0& \a0& \phantom{-}1\\ \a0& \a0& \phantom{-}1& \a0\\ \a0& -1& \a0& \a0\\ -1& \a0& \a0& \a0}\ ,\quad 
  i\MAT{0& \a0& \a0& -1\\ 0& \a0& \phantom{-}1& \a0\\ 0& -1& \a0& \a0\\ 1& \a0& \a0& \a0}
}
and
\SP{
\bar\BSigma_{AB}&=-\MAT{\a0& \phantom{-}1& \a0& \a0\\ -1& \a0& \a0& \a0\\ \a0& \a0& \a0& \phantom{-}1\\ \a0& \a0& -1& \a0}\ ,\quad  
  i\MAT{\a0& \phantom{-}1& \a0& \a0\\ -1& \a0& \a0& \a0\\ \a0& \a0& \a0& -1\\ \a0& \a0& \phantom{-}1& \a0}\ ,\quad  -\MAT{0& \a0& -1& 
        \a0\\ 0& \a0& \a0& \phantom{-}1\\ 1& \a0& \a0& \a0\\ 0& -1& \a0& \a0}\ ,\\  
&\qquad\qquad  i\MAT{0& \a0& -1& \a0\\ 0& \a0& \a0& -1\\ 1& \a0& \a0& \a0\\ 0& \phantom{-}1& \a0& \a0}\ ,\quad  -\MAT{\a0& \a0& \a0& 
        \phantom{-}1\\ \a0& \a0& \phantom{-}1& \a0\\ \a0& -1& \a0& \a0\\ -1& \a0& \a0& \a0}\ ,\quad 
  i\MAT{0& \a0& \a0& -1\\ 0& \a0& \phantom{-}1& \a0\\ 0& -1& \a0& \a0\\ 1& \a0& \a0& \a0}\ .
}
We also define 
\EQ{
\Sigma_{ab}{}^A{}_B=
\tfrac14(\Sigma_a^{AC}\bar\Sigma_{bCB}-\Sigma_b^{AC}\bar\Sigma_{aCB})\ ,\qquad
\bar\Sigma_{abA}{}^B=\tfrac14(\bar\Sigma_{aAC}\Sigma_b^{CB}-
\bar\Sigma_{bAC}\Sigma_a^{CB})\ .
}

\end{document}